\documentclass[11pt]{article}

\usepackage[top=2cm, bottom=2cm, left=2cm, right=2cm]{geometry}

\usepackage{graphicx}
\usepackage{float}
\usepackage{wrapfig}
\usepackage{epstopdf}

\usepackage[latin1]{inputenc}

\usepackage{amsthm,dcolumn,graphicx,multicol,fancyhdr}
\usepackage{latexsym,subfig,graphicx,ifthen}
\usepackage{rotating,amsfonts, color,amssymb,amsmath,amscd,array,enumerate,afterpage}

\numberwithin{equation}{section}

\usepackage{hyperref}
\usepackage{multirow}

\usepackage{enumitem}
\setlist{parsep=3pt,listparindent=\parindent}

\usepackage{chngcntr}
\counterwithout{equation}{section}

\theoremstyle{definition}

\def\subfigure{\subfloat}

\makeatletter\@addtoreset{table}{section}\makeatother
\makeatletter\@addtoreset{figure}{section}\makeatother

\newcommand{\babc}{\renewcommand{\labelenumi}{(\alph{enumi})}\begin{enumerate}}
\newcommand{\eabc}{\end{enumerate}}
\newcommand{\biii}{\renewcommand{\labelenumi}{(\roman{enumi})}\begin{enumerate}}
\newcommand{\eiii}{\end{enumerate}}

\newcommand{\beqn}{\begin{eqnarray*}}
\newcommand{\beq}{\begin{eqnarray}}
\newcommand{\eeqn}{\end{eqnarray*}}
\newcommand{\eeq}{\end{eqnarray}}

\DeclareMathOperator* {\argmax}{ arg\,max}

\newcommand{\ckboldon}[1]{#1}

\newcommand{\ckbold}[1]{%
 \ifthenelse{\isundefined{\ckboldon}}{#1}{ \textbf{#1} }
}

\newcommand{\tr}{\mbox{tr}\,}

\begin{document}
\date{}
\title{Estimating Time-Varying Effective Connectivity in High-Dimensional fMRI Data Using Regime-Switching Factor Models}

\author{
Chee-Ming Ting\footnote{Center for Biomedical Engineering, Universiti Teknologi Malaysia (UTM), 81310 Skudai, Johor, Malaysia;\texttt{cmting@utm.my}},
Hernando Ombao\footnote{Department of Statistics, University of California, Irvine CA 92697, USA; \texttt{hombao@uci.edu}},
S. Balqis Samdin\footnote{Center for Biomedical Engineering, UTM, 81310 Skudai, Johor, Malaysia; \texttt{sbalqis4@live.utm.my}},
Sh-Hussain Salleh\footnote{Center for Biomedical Engineering, UTM, 81310 Skudai, Johor, Malaysia;\texttt{hussain@fke.utm.my}}
}

\maketitle

\begin{abstract}
Recent studies on analyzing dynamic brain connectivity rely on sliding-window analysis or time-varying coefficient models which are unable to capture both smooth and abrupt changes simultaneously. Emerging evidence suggests state-related changes in brain connectivity where dependence structure alternates between a finite number of latent states or regimes. Another challenge is inference of full-brain networks with large number of nodes. We employ a Markov-switching dynamic factor model in which the state-driven time-varying connectivity regimes of high-dimensional fMRI data are characterized by lower-dimensional common latent factors, following a regime-switching process. It enables a reliable, data-adaptive estimation of change-points of connectivity regimes and the massive dependencies associated with each regime. We consider the switching VAR to quantity the dynamic effective connectivity. We propose a three-step estimation procedure: (1) extracting the factors using principal component analysis (PCA) and (2) identifying dynamic connectivity states using the factor-based switching vector autoregressive (VAR) models in a state-space formulation using Kalman filter and expectation-maximization (EM) algorithm, and (3) constructing the high-dimensional connectivity metrics for each state based on subspace estimates. Simulation results show that our proposed estimator outperforms the K-means clustering of time-windowed coefficients, providing more accurate estimation of regime dynamics and connectivity metrics in high-dimensional settings. Applications to analyzing resting-state fMRI data identify dynamic changes in brain states during rest, and reveal distinct directed connectivity patterns and modular organization in resting-state networks across different states.
\end{abstract}

\vspace{5mm}

\section{Introduction}
Most analyses of functional connectivity (FC) using functional magnetic resonance imaging (fMRI) data implicitly assumed that relationships between distinct brain regions are static (stationary) across time. Time-invariant FC metrics such as correlations between fMRI time series are computed over the entire period of recording. Recent years have seen increased interest in investigating dynamic changes in FC patterns over time, often referred to as \textit{dynamic (time-varying) functional connectivity} \cite{Hutchison2013,Calhoun2014}. Several studies have reported temporal fluctuations in FC at time-scales of seconds to minutes, in both strength and directionality of the connections, even during resting state \cite{Chang2010a,Allen2012,Leonardi2013,Hutchison2013b,Zalesky2014}.

The simplest and most common approach to examining the dynamic behavior in connectivity is the sliding-window correlation, which involves computing locally stationary correlations over consecutive windowed short-time segments of data to produce time-varying FC metrics \cite{Chang2010a,Allen2012,Hutchison2013,Zalesky2014}. However, this approach is limited by the choice of optimal window length: a long window has low statistical power to detect abrupt and highly localized changes, while a short window produces noisy estimates for smooth changes. An alternative strategy is the model-based approach, which can provide a unified, parsimonious framework to characterize the dynamic connectivity structure based on the time-dependent model parameters. For example, the time-varying multivariate volatility models \cite{Lindquist2014} and the time-varying vector autoregressive (VAR) models \cite{Havlicek2010,Samdin2015} have been used to capture effectively instantaneous temporal changes in fMRI-based functional and effective connectivity (a more specific cross-dependence with directionality, in a sense that it measures the causal influence of one brain region on another).

Recent evidence from fMRI studies suggested state-related types of dynamics in FC: time-varying but reoccurring connectivity patterns which switches according to a few discrete underlying quasi-stable brain states (regimes). This non-stationarity is characterized by rapid transitions between regimes and smooth changes within a regime. Various analytical approaches have been used to identify these replicable dynamic `connectivity states'. These include K-mean clustering of the windowed correlations \cite{Allen2012,Hutchison2013}, which, however, ignores information about the temporal order of the dynamics, hidden Markov models producing the state-time alignments \cite{Baker2014}, or algorithms to detect change points in connectivity \cite{Cribben2012,Jeong2016}. However, these studies focused on evaluating the un-directed connectivity. Our recent work \cite{Balqis2016} proposed a more general method based on the switching VAR (SVAR) models to infer dynamic states of effective connectivity in fMRI and EEG data. The next challenge is to estimate the high-dimensional connectivity states for a large number of brain regions, where traditional analyses based on pair-wise correlations or using average signals from parcellated regions of interest (ROIs) might produce sub-optimal results.

In this paper, we propose a new approach based on regime-switching factor models for estimating temporal changes in effective connectivity states in high-dimensional fMRI data for a whole-brain network analysis. Precisely, our approach is to first employ a factor analysis model to characterize the large fMRI data via a small number of common, latent (unobservable) factor components, and then identify the dynamic connectivity regimes based on these low-dimensional summary signals. We develop a non-stationary factor model which takes into account the time-variation of the underlying serial cross-correlation structure of the high-dimensional data, by introducing regime-switching in the factor dynamics, By specifying the factors to evolve as a Markov-switching VAR process, we derive a factor SVAR model for the observation space, which is an extension of the SVAR model used in \cite{Balqis2016} to the large-dimensional case. Such formulation implies projection of the high-dimensional directed connectivity matrix onto a lower-dimensional subspace (small VAR coefficient matrix of factors, spanned by the factor loadings), and thus allows us to capture the changes in connectivity regimes in these subspaces driven by the few factors. It enables a reliable and computationally-efficient estimation of the regime change-points and the massive dependence measures associated with each regime.

We develop a three-step estimation procedure. The first step is initial estimation of connectivity subspace shared across regimes based on a stationary factor model. The number of factors and a common factor loading (specifying the dimension and the span of the subspace) are estimated by applying the principal component analysis (PCA) on the data. The second step is the dynamic regime segmentation based on the factor SVAR model formulated in a state-space form. The change-points between connectivity states are identified via switching Kalman filter and switching Kalman smoother (SKF and SKS). Simultaneously, the regime-dependent parameters of the latent switching factor process are updated by the expectation-maximization (EM) algorithm, with the common factor loadings initialized and fixed using the estimates from the first step. Then, the fMRI signals are partitioned according to the estimated states, and fitted with a separate factor model for each regime to obtain regime-dependent factor loadings. The third step is estimation of within-regime connectivity metrics, where the estimates of the high-dimensional VAR connectivity graph/matrix for each state are constructed from the low-dimensional factor subspace parameters estimated from the previous two steps. We evaluated the performance of our method via simulations by comparing with the K-mean clustering approach. Application to the resting-state fMRI data reveals switching states of the resting state connectivity networks, with the modular organization changes across different states.

\section{Regime-switching Factor VAR Models}
In this section, we first describe the stationary factor model with an autoregressive factor process. Then, we introduce a non-stationary generalization with regime-switching in the factor dynamics.

\subsection{The Factor Model}
Let ${\bf y}_t = [y_{1t}, \ldots, y_{Nt}]'$ be a $N \times 1$ observed vector of non-stationary time series of fMRI at time points $t=1, \ldots, T$. The cross-section dimension of the time series $N$ can be comparable to or even larger than the sample size $T$ (or length of time series). We suppose the high-dimensional time series is driven by a small number of latent factors. Specifically, we consider a factor model defined by
\begin{equation} \label{Eq:fm}
{\bf y}_t = {\bf Q} {\bf f}_t + {\boldsymbol{\epsilon}}_t
\end{equation}
where ${\bf f}_t = [f_{t1}, \ldots, f_{tr}]'$ is a $r \times 1$ vector of unobserved common factors with mean zero and covariance matrix $\boldsymbol{\Sigma}_{\bf f}$, ${\bf Q} = [{\bf q}_1 , \ldots, {\bf q}_r]$ is a $N \times r$ constant factor loading matrix assumed to be orthonormal, i.e. ${\bf Q}' {\bf Q} = {\bf I}_{r}$ where ${\bf I}_{r}$ denotes a $r \times r$ identity matrix, $r$ is the number of factors satisfying $r << N$, and ${\boldsymbol{\epsilon}}_t = [\epsilon_{t1}, \ldots, \epsilon_{tN}]'$ is $N \times 1$ vector of noise components with mean zero and covariance matrix $\boldsymbol{\Sigma}_{\boldsymbol{\epsilon}} = {\mbox{diag}}({\sigma}_{\epsilon_{1}}^2, \ldots, {\sigma}_{\epsilon_{N}}^2)$, assuming the error terms are cross-sectionally uncorrelated. The model captures the correlation between the time series via the mixing of some common factors ${\bf f}_t$ by ${\bf Q}$. The model (\ref{Eq:fm}) allows for dimension-reduction in the sense that the serial and cross-correlation in the high-dimensional observational process $\{{\bf y}_t\}$ is driven by the much lower-dimensional factor process $\{{\bf f}_t\}$ and mixing matrix ${\bf Q}$.

The evolution of the latent factor dynamics in $\{{\bf f}_t\}$ can be modeled by a stationary vector autoregressive (VAR) process of order $P$, VAR($P$)
\begin{equation} \label{Eq:VAR-f}
{\bf f}_t = {\bf \Phi}_{\bf f}(1) {\bf f}_{t-1} + \ldots + {\bf \Phi}_{\bf f}(P) {\bf f}_{t-P} + {\boldsymbol \eta}_t, \ \ \ {\boldsymbol \eta}_t \sim N({\bf 0}, \boldsymbol{\Sigma}_{\boldsymbol \eta})
\end{equation}
where ${\bf \Phi}_{\bf f}(\ell)$ is the $r \times r$ AR coefficients matrix at lag $\ell$ for $l = 1, \ldots, P$ and ${\boldsymbol \eta}_t$ is a $r \times 1$ Gaussian white noise process with mean zero and covariance matrix $\boldsymbol{\Sigma}_{\boldsymbol \eta}$. Both processes $\{{\bf y}_t\}$ and $\{{\bf f}_t\}$ are non-stationary where the factor loadings ${\bf Q}$ and the AR coefficients matrices for factors ${\bf \Phi}_{\bf f}(\ell)$ are time-constant. \\

\textit{\underline{Factor VAR Model}}: The temporal inter-dependence in the high-dimensional observation process ${\bf y}(t)$ can be characterized by the much lower-dimensional VAR process of ${\bf f}_t$ in (\ref{Eq:VAR-f}). This forms the basic idea of our recent works \cite{Ting2014,Wang2016}, where we developed a factor-based VAR (f-VAR) model for the observations ${\bf y}(t)$ by substituting (\ref{Eq:VAR-f}) into (\ref{Eq:fm}) and assuming ${\boldsymbol{\epsilon}}_t$ approximately zero, which gives
\begin{equation}\label{Eq:VAR-y}
{\bf y}_t = \sum \limits _{\ell = 1}^{P} {\bf \Phi}_{\bf y}(\ell) {\bf y}_{t-\ell} + {\bf Q} {\boldsymbol \eta}_t
\end{equation}
where ${\bf \Phi}_{\bf y}(\ell) = {\bf Q} {\bf \Phi}_{\bf f}(\ell) {\bf Q}'$ are high-dimensional $N \times N$ coefficients matrices for ${\bf y}_t$, an orthogonal projection of the smaller matrices ${\bf \Phi}_{\bf f}(\ell)$ on to lower-dimensional subspace spanned by the columns of ${\bf Q}$. It provides a low-rank approximation for the dependence structure in ${\bf y}(t)$. The model subspace can be learned by using the principal component analysis (PCA) where the estimator for ${\bf Q}$ are defined by eigenvectors corresponding to the $r$ largest eigenvalues of the sample covariance matrix of ${\bf y}(t)$. It leads to substantially improved consistency and computational efficiency in estimating VAR models under high-dimensional settings, compared to the traditional least-squares estimator as shown in \cite{Ting2014}. 

The coefficients matrix ${\bf \Phi}_{\bf y}(\ell)$ can quantify directed interactions in a network with large number of nodes (e.g. a large-scale network of brain regions) at time lag $\ell$. There exists a directed influence in the Granger-causality sense with direction from node $j$ to node $i$ for any connection strength $\vert {\bf \Phi}_{ij} \vert > 0$, where ${\bf \Phi}_{ij}$ is the ($i,j$)-th element of ${\bf \Phi}_{\bf y}$. When applied to identify effective brain connectivity networks with a large number of nodes from resting-state fMRI data \cite{Ting2014}, the estimates provided more reliable interpretation and capable of revealing the modular, hierarchical structure of the brain networks during rest, by varying the subspace dimension $r$.

\subsection{Regime-switching in Factor Dynamics}

\textit{\underline{Factor Model with Regime-Switching}}: We now generalize the stationary factor model in (\ref{Eq:fm}) to allow for time-variation in the serial interdependence structure of the latent factors, by introducing regime-switching in the coefficient matrices of the VAR factor specification in (\ref{Eq:VAR-f}). In this respect, we propose a non-stationary factor model with regime-switching factor dynamics. More precisely, we assume the factor loadings remain stationary but the factors to follow a Markov-switching VAR (SVAR) process of order $P$, SVAR($P$). This class of models has been applied for modeling of econometric data \cite{Krolzig2013}. The SVAR is a quasi-stationary model consisting of a set of $K$ independent VAR models, each indexed by a hidden random indicator $S_t$
\begin{equation} \label{Eq:SVAR-f}
{\bf f}_t = \sum_{\ell=1}^{P} {\bf \Phi}_{\bf f}^{[S_t]}(\ell) {\bf f}_{t-\ell} + {\boldsymbol \eta}_t, \ \ \ {\boldsymbol \eta}_t \sim N({\bf 0}, \boldsymbol{\Sigma}_{\boldsymbol \eta}^{[S_t]})
\end{equation}
here $\{S_t \in \{j = 1, \ldots, K\}, \ t=1, \ldots, T\}$ is a sequence of state/regime variables, which is time-dependent and take values in a discrete space $j = 1, \ldots, K$; and $\{{\bf \Phi}_{\bf f}^{[j]}(\ell), \ell=1, \ldots, P\}$ are coefficient matrices for state $j$. This is a generalized version of (\ref{Eq:VAR-f}) which allows for structural changes in the VAR coefficients. The AR coefficients matrices ${\bf \Phi}_{\bf f}^{[S_t]}(\ell)$ are piecewise constant function of the discrete state $S_t$, i.e. constant within time-blocks belong to a same regime but change across different regimes. This renders the factor process piecewise stationary, a special form of departure from stationarity. However, the proposed model differs from the classical piecewise constant processes  (e.g., piecewise VAR processes) primarily because in the classical piecewise processes the future blocks are not at all related to previous blocks. Our proposed model permits recurring regimes where future blocks could be related to past blocks if they were both indexed by the same state. This has important implications in estimation and inference because we can pool together different time-blocks of the same regime (that are indexed by the same state) thus producing more accurate and more efficient estimates

We assume that $S_t$ follows a $K$-state first-order Markovian process with a $K \times K$ transition matrix $Z = [z_{ij}], 1 \leq i,j \leq K$ where
\begin{equation}
z_{ij} = P(S_{t}=j \vert S_{t-1}=i) \label{eq5}
\end{equation}
denotes the probability of transition from state $i$ at time $t-1$ to state $j$ at $t$. Only one latent process (and hence only one VAR process) is ``active" (or turned on) at each time point $t$. The remaining latent processes are turned off. This allows recurring changes in the temporal interdependence structure of the factors as characterized by ${\bf \Phi}_{\bf f}^{[S_t]}$, which switches over time between the finite number of regimes, according to the regime indicator $S_t$ at time $t$. Compared to using a switching VAR model directly on ${\bf y}_t$, the specification of Equation (\ref{Eq:SVAR-f}) allows us to detect the change-points of the high-dimensional dependence structure based on a small number of factor series. We denote the model parameters by $\boldsymbol{\theta}= \{ \boldsymbol{\theta}_{j} = ({\bf \Phi}_{\bf f}^{[j]},\boldsymbol{\Sigma}_{\boldsymbol \eta}^{[j]}): j \in \{1, \ldots, K \} \}$ which are assumed unknown and to be estimated.\\

\textit{\underline{Factor Switching-VAR Model}}: We shall derive a high-dimensional switching VAR model from the non-stationary factor model with a regime-switching autoregressive factor process as defined by Equation (\ref{Eq:fm}) and (\ref{Eq:SVAR-f}). The regime-switching in the high-dimensional interdependence structure of observations $\{{\bf y}_t\}$ can be driven by that of the lower-dimensional SVAR factor process in (\ref{Eq:SVAR-f}). Substituting (\ref{Eq:SVAR-f}) into (\ref{Eq:fm}) yields
\begin{eqnarray}
{\bf y}_t &=&  {\bf Q}\Big ( \sum_{\ell=1}^{P} {\bf \Phi}_{\bf f}^{[S_t]}(\ell) {\bf f}_{t-\ell} + {\boldsymbol \eta}_t \Big ) \\
&=& \sum \limits _{\ell = 1}^{P} {\bf Q} {\bf \Phi}_{\bf f}^{[S_t]}(\ell) {\bf Q}' {\bf Q} {\bf f}_{t-\ell} + {\bf Q} {\boldsymbol \eta}_t \\
&=& \sum \limits _{\ell = 1}^{P} {\bf Q} {\bf \Phi}_{\bf f}^{[S_t]}(\ell) {\bf Q}' {\bf y}_{t-\ell}  + {\bf Q} {\boldsymbol \eta}_t.
\end{eqnarray}
Finally, we have a factor-based Markov-switching VAR (f-SVAR) for ${\bf y}_t$
\begin{equation}\label{Eq:SVAR-y}
{\bf y}_t = \sum \limits _{\ell = 1}^{P} {\bf \Phi}_{\bf y}^{[S_t]}(\ell) {\bf y}_{t-\ell} + {\boldsymbol \nu}_t
\end{equation}
where ${\bf \Phi}_{\bf y}^{[S_t]}(\ell) = {\bf Q} {\bf \Phi}_{\bf f}^{[S_t]}(\ell) {\bf Q}'$ and ${\boldsymbol \nu}_t = {\bf Q} {\boldsymbol \eta}_t$. The model is a nonstationary generalization of the factor VAR model in (\ref{Eq:VAR-y}), by allowing a regime-switching in the coefficient parameters. It provides a tool to capture the regime-switching in the large $N$-dimensional serial inter-dependence structure via a low-dimensional space. The model can quantify dynamics of a large-scale directed network with state-dependent changes in the network structure, i.e. switching according to distinct states. It enable the detection of the temporal change points in the network dependency structure, as well as estimation of the directed dependencies between massive number of nodes associated with each state. \\

\textit{\underline{State-Space Formulation}}: We propose a state-space representation for the factor model with regime-switching factors, to enable sequential estimation in time of the latent factors and the switching states. The latent switching VAR factor process (\ref{Eq:SVAR-f}) forms the state-equation which is projected to the high-dimensional space using the factor model (\ref{Eq:fm}) with an error as the observation equation. Defining the dynamic factor structure ${\bf F}_t = [{\bf f}'_t, {\bf f}'_{t-1}, \ldots, {\bf f}'_{t-P+1}]'$ as state vector, the model (\ref{Eq:fm}) and (\ref{Eq:SVAR-f}) are formulated in a switching linear Gaussian state-space form \cite{Kim1994}
\begin{eqnarray}
{\bf F}_t & = & {\bf A}_{\bf F}^{[S_t]}{\bf F}_{t-1} + \mathbf{w}_t 			\label{Eq:ssm-f} \\
{\bf y}_t & = & {\bf H}{\bf F}_t + {\boldsymbol{\epsilon}}_t 	\label{Eq:ssm-obs}
\end{eqnarray}
The SVAR($P$) factor process (\ref{Eq:SVAR-f}) is re-written in a SVAR($1$) form of (\ref{Eq:ssm-f}) in the state equation, where $\mathbf{w}_t = [{\boldsymbol{\eta}}'_t, {\bf 0}', \ldots, {\bf 0}']'$ is $rP \times 1$ state noise, and ${\bf A}_{\bf F}^{[S_t]}$ is a $rP \times rP$ state transition matrix switching with the state variables $S_t$, and of the form
\[
{\bf A}_{\bf F}^{[S_t]} =
\left(
  \begin{array}{ccccc}
    {\bf \Phi}_{\bf f}^{[S_t]}(1) & {\bf \Phi}_{\bf f}^{[S_t]}(2) & \ldots & {\bf \Phi}_{\bf f}^{[S_t]}(P-1) & {\bf \Phi}_{\bf f}^{[S_t]}(P) \\
    {\bf I}_r & {\bf 0} & \ldots & {\bf 0} & {\bf 0} \\
    {\bf 0} & {\bf I}_r  & \ldots & {\bf 0} & {\bf 0} \\
		\vdots &   & \ddots  &  & \vdots \\
		{\bf 0} & {\bf 0} & \ldots & {\bf I}_r & {\bf 0} \\
  \end{array}
\right).
\]
The matrix ${\bf A}_{\bf F}^{[S_t]}$ describes the directed connectivity that varies across states. The unobserved SVAR($P$) dynamic factors ${\bf F}_t$ now follows a (higher dimensional) latent SVAR(1) process. A noisy version of factor model is re-formulated from (\ref{Eq:fm}) as in the observation equation (\ref{Eq:ssm-obs}) by introducing an idiosyncratic noise $\boldsymbol{\epsilon}_t$, and with a $r \times rP$ mapping matrix $\mathbf{H} = [{\bf Q}, {\bf 0}, \ldots, {\bf 0}]$. We assume both $\{ {\boldsymbol{\epsilon}}_t \}$ and $\{ \mathbf{w}_t \}$ are white Gaussian noise, $\boldsymbol{\epsilon}_t \sim N(\mathbf{0},\boldsymbol{\Sigma}_{\boldsymbol \epsilon})$ and $\mathbf{w}_t \sim N(\mathbf{0},\boldsymbol{\Sigma}_{\bf w}^{[S_t]})$, with a time-constant state noise covariance matrices $\boldsymbol{\Sigma}_{\boldsymbol \epsilon}$ and the $\boldsymbol{\Sigma}_{\bf w}^{[S_t]}$ switching with $S_t$. Both the factor loadings ${\bf Q}$ and the noise covariance matrix in the observation equation are assumed to be regime-invariant and shared across regimes. The processes $\{{\bf f}_t\}$ and $\{{\boldsymbol{\epsilon}}_t\}$ are uncorrelated. Instead of hard state assignment for each time-point $t$, we can evaluate the probability of activation for each state, $P(S_t=j|{\bf y}_{1:T})$, which is termed ``soft-alignment''. We denote all model parameters from each of the states as $\boldsymbol{\Theta}= \{ \boldsymbol{\Theta}_{j} = ({\bf A}_{\bf F}^{[j]},\boldsymbol{\Sigma}_{\bf w}^{[j]}): j \in \{1, \ldots, K \} \}$.

\section{Estimation}
We develop a three-step procedure for efficiently estimating the dynamic connectivity states in the high-dimensional fMRI data based on the proposed non-stationary factor model with regime-switching. In the first step, we explore the connectivity subspace assumed as common and shared across regimes, by fitting a stationary factor model (\ref{Eq:fm}) to the entire fMRI time series. We apply the method of PCA to estimate the factor loadings ${\bf Q}$ and the latent factors ${\bf f}_t$, and the Bayesian information criterion (BIC) of \cite{Bai2002} to select the optimal number of factors. In the second step, we perform connectivity regime segmentation in the low-dimensional subspace relying on a factor model with Markov-switching VAR factor process (\ref{Eq:SVAR-f}). Based on the state-space representation (\ref{Eq:ssm-f})-(\ref{Eq:ssm-obs}), the latent factor process can be jointly estimated conditioned on the observational factor model. The temporal change-points of the regimes can be detected via the estimated state sequence $\{\widehat{S}_t\}$ by the SKF and SKS, and the factor VAR coefficient matrix ${\bf \Phi}_{\bf f}^{[j]}$ for each regime is updated iteratively using the EM algorithm. In the third step, we estimate the regime-dependent high-dimensional connectivity matrix ${\bf \Phi}_{\bf y}^{[j]}$ for the observation space using the estimated subspace parameters from the first two steps.

\subsection{Step 1: Estimation of a Common Factor Model}
PCA is a common approach to estimating approximate factor model based on the eigen-decomposition of sample covariance matrix \cite{Bai2003,StockWatson2002}. Let ${\bf V}_{1}$, $\ldots$, ${\bf V}_{N}$ be $N$ orthonormal eigenvectors corresponding to the eignevalues of the $N \times N$ sample covariance matrix ${\bf S}_{\bf y} = \sum_{t=1}^{T} {\bf y}_t {\bf y}'_t$, in a decreasing order such that $\widehat{\lambda}_{1} \geq \ldots \geq \widehat{\lambda}_{N} >0$. The PCA estimator of the loadings $\widehat{\bf Q} = \left[ {\bf V}_{1}, \ldots, {\bf V}_{r} \right]$ is defined by a matrix whose columns are the $r$ orthonormal eigenvectors corresponding to the largest $r$ eignevalues, and the factors can be estimated by $\widehat{\bf f}_t = \widehat{\bf Q}'{\bf y}_t$. \cite{Bai2003} has showed that the PCA estimators are consistent and asymptotically normal, under settings of large $N$ and large $T$. Besides, the estimates can be computed efficiently even under situations when $N < T$ (on the small $T \times T$ temporal covariance matrix instead of the huge $N \times N$ spatial sample covariance ${\bf S}_{\bf y}$). We can compute the noise covariance estimator as $\widehat{\boldsymbol{\Sigma}}_{\boldsymbol{\epsilon}} = \sum_{t=1}^{T} \widehat{\boldsymbol{\epsilon}}_t \widehat{\boldsymbol{\epsilon}}'_t$ based on the residuals $\widehat{\boldsymbol{\epsilon}}_t = {\bf y}_t - \widehat{\bf Q}\widehat{\bf f}_t$ from the fitted factor model. We fit an VAR model (\ref{Eq:VAR-f}) to the estimated factors $\{ \widehat{\bf f}_t \}$ by the least-squares (LS) method, and obtain the AR coefficient estimates $\widehat{\bf \Phi}_{\bf f}(\ell)$. For PCA estimation, the number of factors can be determined by model selection using BIC
\begin{equation}\label{Eq:BIC}
\hat{r} = \argmax_{\{1, \ldots, L_r\}} \left\{ \ln \left(\frac{1}{NT} \sum_{t=1}^{T} \|\widehat{\boldsymbol{\epsilon}}_t(r)\|_2^2\right) + r \left(\frac{N+T}{NT}\right) \ln\left(\frac{NT}{N+T}\right) \right\}
\end{equation}
where $\|{\bf x}\|$ denotes the Euclidean norm of a vector ${\bf x}$ and $L_r$ is a bounded integer such that $r \leq L_r$.

\subsection{Step 2: Estimation of Regime-switching Factor Model}
Based on the state-space formulation (\ref{Eq:ssm-f})-(\ref{Eq:ssm-obs}), the objective is to extract the underlying states $\{\widehat{S}_t\}$, and to estimate the unknown coefficient matrix ${\bf \Phi}_{\bf f}^{[j]}$ and factor signals in ${\bf F}_t = [{\bf f}'_t, {\bf f}'_{t-1}, \ldots, {\bf f}'_{t-P+1}]'$ of the latent Markov-SVAR factor process given observations $\widehat{\bf y}_t, t = 1, \ldots T$. \\

\textit{\underline{Filtering and Smoothing}}: The inference of $S_t$ and ${\bf F}_t$ involve computing, sequentially in time, the filtered probabilities $Pr(S_t| {\bf y}_{1:t})$ and the filtered densities $p({\bf F}_{t}|{\bf y}_{1:t})$, given the available signal observations up to time $t$, ${\bf y}_{1:t} = \{ {\bf y}_{1}, \ldots, {\bf y}_{t} \}$, and the more accurate smoothed probabilities $P(S_t|{\bf y}_{1:T})$ and densities $p({\bf F}_{t}|{\bf y}_{1:T})$ given the available entire set of observations ${\bf y}_{1:T} = \{ {\bf y}_{1}, \ldots, {\bf y}_{T}\}$. We estimate the filtered and smoothed densities of ${\bf F}_{t}$ given state $j$ at time $t$, by the KF and the KS, respectively
\begin{eqnarray}
{\bf F}_{t|t}^{j} & = & \text{E}({\bf F}_t|{\bf y}_{1:t},S_{t}=j)		 						     \label{eq11} 	\\
				 V_{t|t}^{j} & = & \text{Cov}({\bf F}_t|{\bf y}_{1:t},S_{t}=j)		 					     \label{eq12}  \\
\mathbf{X}_{t|T}^{j} & = & \text{E}({\bf F}_t|{\bf y}_{1:T},S_{t}=j)		 						     \label{eq13} \\
				 V_{t|T}^{j} & = & \text{Cov}({\bf F}_t|{\bf y}_{1:T},S_{t}=j) 		 					     \label{eq14} \\
		 V_{t,t-1|T}^{j} & = & \text{Cov}({\bf F}_t,{\bf F}_{t-1}|{\bf y}_{1:T},S_{t}=j)    \label{eq15}
\end{eqnarray}
where ${\bf F}_{t|t}^{j}$ and $V_{t|t}^{j}$ are mean and covariance of the filtered density $p({\bf F}_t|{\bf y}_{1:t},S_{t}=j)$, ${\bf F}_{t|T}^{j}$ and $V_{t|T}^{j}$ are mean and covariance of the smoothed density $p({\bf F}_t|{\bf y}_{1:T},S_{t}=j)$ given state $j$ at time $t$, and $V_{t,t-1|T}^{j}$ is the cross-variance of joint density $p({\bf F}_t, {\bf F}_{t-1}|{\bf y}_{1:T},S_{t}=j)$. The estimates of filtered and smoothed state occupancy probability of being state $j$ at time $t$ are also computed as
\begin{eqnarray}
\mathbf{M}_{t|t}^{j} & = & P(S_t=j|{\bf y}_{1:t}) \label{eq16} \\																		
\mathbf{M}_{t|T}^{j} & = & P(S_t=j|{\bf y}_{1:T}) \label{eq17}
\end{eqnarray}

\textit{\underline{EM Estimation}}: The estimates of the factor-subspace dynamic parameters in ${\bf A}_{\bf F}^{[j]}$ and $\boldsymbol{\Sigma}_{\bf w}^{[j]}$ can be obtained by the maximum likelihood (ML) method by maximizing the log-likelihood ${L} = \log p({\bf y}_{1:T}|\boldsymbol{\Theta})$ with respect to each parameter. Here, we use the EM algorithm for the switching state-space model suggested by \cite{Murphy1998}. In the expectation step (E-step), the sufficient
statistics are obtained from the smoothed estimates
\begin{eqnarray}
P_t 		  & = & \text{E}({\bf F}_t{\bf F}_t'|{\bf y}_{1:T}) = V_{t|T}+{\bf F}_{t|T}{\bf F}_{t|T}' 						\label{eq18} 	\\
P_{t,t-1} & = & \text{E}({\bf F}_t{\bf F}_{t-1}'|{\bf y}_{1:T}) = V_{t,t-1|T}+{\bf F}_{t|T}{\bf F}_{t-1|T}' 	\label{eq19} 	
\end{eqnarray}
where ${\bf F}_{t|T}$, $V_{t|T}$ and $V_{t,t-1|T}$ are quantities of the smoothed densities $p({\bf F}_{t}|{\bf y}_{1:T})$ and $p({\bf F}_{t}, {\bf F}_{t-1}|{\bf y}_{1:T})$, corresponding to (\ref{eq13}) to (\ref{eq15}) by marginalizing out the state variable $j$ of the $p({\bf F}_{t}|{\bf y}_{1:T},S_{t}=j)$ and $p({\bf F}_{t}, {\bf F}_{t-1}|{\bf y}_{1:T},S_{t}=j)$ using Gaussian approximation. We retain the terms switching KF (SKF) and switching KS (SKS) to refer to KF/KS approach to estimating state parameters of the SVAR model, as in \cite{Murphy1998}.

In the maximization step (M-step), the estimates of the model parameters for regime $j$ are updated as follows
\begin{eqnarray}
\widehat{\bf A}_{\bf F}^{[j]} & = & \left(\sum_{t=2}^T W_t^j P_{t, t-1}\right) \left(\sum_{t=2}^TW_t^j P_{t-1} \right)^{-1} \label{eq20} \\
\boldsymbol{\Sigma}_{\bf w}^{[j]} & = & \left(\dfrac{1}{\sum_{t=2}^T W_t^j}\right) \left(\sum_{t=2}^T W_t^j P_{t} - \widehat{\bf A}_{\bf F}^{[j]}\sum_{t=2}^TW_t^j P_{t,t-1}'\right) \label{eq21} \\
\widehat{z}_{ij} & = & \dfrac{\sum_{t=2}^T P(S_{t-1}=j, S_t=i|{\bf y}_{1:T})}{\sum_{t=1}^{T-1} W_t^j} \label{eq22}
\end{eqnarray}
where the weights $W_t^j = \mathbf{M}_{t|T}^{j}$ are computed from the smoothing step. The model parameters are iteratively until some convergence criteria are satisfied, to produce the ML estimates $\boldsymbol{\Theta}^{*}$. We used randomized initial estimates for entries of $\widehat{\bf A}_{\bf F}^{[j]}$. The factor loading matrix ${\bf Q}$ in $\mathbf{H}$ and the noise covariance $\boldsymbol{\Sigma}_{\boldsymbol \epsilon}^{[j]}$ which are assumed common to all regimes, remain fixed with the PCA estimates from Step 1, and not updated by the EM algorithm. Note that here the regime estimation is done based on the state equation of low-dimensional factors. This will lead to substantial computational reduction, and improve the identifiability of the individual subspace parameter estimators. \\

\textit{\underline{Regime Segmentation}}: Given the EM-estimated model parameters $\boldsymbol{\Theta}^{*}$, the reliminary temporal regime segmentation $\widehat{\bf \Phi}_{\bf f}^{[\widehat{S}_t]}$ in the subspace is defined by the latent state sequence estimated using the SKF, $\widehat{S}^{\text{SKF}}_t = \argmax_{j} P(S_t=j|{\bf y}_{1:t})$ in (\ref{eq16}) which indicates the most likely active state for each time point. This is then further refined by the SKS, $\widehat{S}^{\text{SKS}}_t = \argmax_{j} P(S_t=j|{\bf y}_{1:T})$ in (\ref{eq17}) based on both the past and future observations. We can also utilize this state-time alignment provided in $\widehat{S}^{\text{SKF}}_t$ and $\widehat{S}^{\text{SKS}}_t$ to partition the observed fMRI signals into their corresponding states, and the time-segments of each regime is then fitted with a separate stationary factor model to derive state-dependent estimators, as described in the next step.

\subsection{Step 3: Estimation of Regime-dependent Connectivity Matrices}
We investigate two different schemes for constructing the estimators for the high-dimensional VAR-based connectivity matrix or graph for each state $\widehat{\bf \Phi}_{\bf y}^{[j]}$, by plugging in the subspace parameter estimators obtained in the first two steps: (1) Coupled SVAR estimator (with common factor loadings) $\widehat{\bf \Phi}_{\bf y}^{[j]}(\ell) = \widehat{\bf Q} \widehat{\bf \Phi}_{\bf f}^{[j]*}(\ell) \widehat{\bf Q}'$, by substituting in the f-SVAR model in (\ref{Eq:SVAR-y}) with the EM estimate $\widehat{\bf \Phi}_{\bf f}^{[j]*}(\ell)$ from Step 2 and the PCA estimate $\widehat{\bf Q}$ from Step 1. Note that conditioned on a common factor loading $\widehat{\bf Q}$, the factor coefficient matrices $\widehat{\bf \Phi}_{\bf f}^{[j]*}(\ell)$ of all regimes are jointly estimated by the EM, weighted at each state by the smoothed state occupancy probability $P(S_t=j|{\bf y}_{1:T})$ in (\ref{eq17}). (2) Decoupled SVAR estimator (with state-dependent factor loadings) $\widetilde{\bf \Phi}_{\bf y}^{[j]}(\ell) = \widetilde{\bf Q}^{[j]} \widetilde{\bf \Phi}_{\bf f}^{[j]}(\ell) \widetilde{\bf Q}'^{[j]}$ by substituting in a separate f-VAR model in (\ref{Eq:VAR-y}) for each state. ($\widetilde{\bf \Phi}_{\bf f}^{[j]}$, $\widetilde{\bf Q}$) are PCA estimates by fitting distinct stationary factor models (\ref{Eq:fm}) separately to each of the regime time-courses, derived from the SKS segmentation in Step 2. The limiting distribution of the factor-VAR estimator has been derived in \cite{Ting2014} (Theorem 2). For ease of exposition, we drop the state index $j$ and focus on the VAR(1). The subspace estimator $\widetilde{\bf b} = vec(\widetilde{\bf \Phi}_{\bf y}) = vec(\widetilde{\bf Q} \widetilde{\bf \Phi}_{\bf f} \widetilde{\bf Q}')$ has an asymptotic normal distribution as $T \rightarrow \infty$
\begin{equation}\label{Eq:Asym-fVAR}
\sqrt{T} (\widetilde{\bf b} - {\bf b}) \stackrel{D}{\rightarrow} N({\bf 0}, {\bf G})
\end{equation}
where ${\bf G} = ({\bf Q}\boldsymbol{\Sigma}_{\boldsymbol \eta}{\bf Q}')\otimes({\bf Q}\boldsymbol{\Gamma}_{\bf f}{\bf Q}')$ with ${\Gamma}_{\bf f} = cov({\bf f}_t)$ and $\otimes$ denotes the Kronecker product. By replacing with the PCA estimates, the covariance matrix of the estimator can be estimated, defined by $\widehat{\bf G}$. Based on this, we test the significance of each subspace VAR coefficient in $\widetilde{\bf b}$ as being different from zero, with $H_0:b_k = 0$ against $H_1:b_k \neq 0$, where $b_k$ is $k$-th element of ${\bf b}$. The test statistic is approximately distributed as $t_k = \widehat{b}_k / {\sqrt{\widehat{\bf G}_{kk}/T}} \sim N(0,1)$ when $T$ is sufficiently large, where $\widehat{\bf G}_{kk}$ is $k$-th diagonal entry of $\widehat{\bf G}$. A coefficient is significant if the $p-value<\alpha/D$ with $\alpha$ the significance level and $D = N^2$ the number of tested coefficients, implying corrections for multiple testing by Bonferroni method. 

\section{Simulations}

In this section, we evaluate the numerical performance of the proposed factor-SVAR model-based estimators in identifying state-dependent changes in large-scale directed connectivity networks through simulations. The objective is to measure the ability of our estimation procedures in (1.) detecting the change-points of connectivity regimes via the estimated state sequence, and (2) estimating the high-dimensional directed connectivity matrix or graph between nodes for each regime. \\

\textit{\underline{Data Generation}}: We generated data from a regime-switching VAR(1) model with with $K = 2$ states, with different coefficient matrix of the independent VAR for each state to characterize distinct connectivity patterns. To emulate the modular connectivity network structure, we assume a block-diagonal VAR coefficient matrix, formed by $10\times10$ dimensional non-zero sub-blocks along the main diagonal. Each sub-block represents the directed connectivity in a sub-network of 10 nodes. The entries of the sub-blocks were randomly drawn from a uniform distribution. Here, we set the two state-dependent coefficient matrices with distinct structure as ${\bf \Phi}_{\bf y}^{[1]}: a_{ij} \sim \mbox{U}[-0.4~0.4]$ and ${\bf \Phi}_{\bf y}^{[2]}: a_{ij} \sim \mbox{U}[-0.2~0.2]$, for $i$ and $j$ in the same block. The entries of the off-diagonal blocks are zero. We set the same noise covariance matrix for both state $\boldsymbol{\Sigma}_{\boldsymbol \eta} = 0.5{\bf I}$. 

Locally-stationary time-series data with piece-wise stable connectivity structure over time, were obtained by concatenating the two VAR processes simulated independently. The simulated data consists of four time-blocks each from a VAR and of fixed length $T_B=50$ (total length of $T=200$), with 3 change points at times $t=50$, $t=100$ and $t=150$. The sample size available for the VAR model of each state is only $T=100$. To emulate the state-dependent recurring changes in the VAR connectivity structure, the successive time-blocks were generated according to the distinct coefficient matrices in a cyclic manner, alternating between the two connectivity states, following procedure in \cite{Monti2014} for functional connectivity. Thus, the state labels and the corresponding state-dependent VAR coefficient matrices for each time points are considered known and used as ground-truth for evaluation, i.e. ($S_t=1$, ${\bf \Phi}_{\bf y}^{[1]}$) for $t=1, \ldots, 50$ and $t=101, \ldots, 150$; ($S_t=2$, ${\bf \Phi}_{\bf y}^{[2]}$) for $t=51, \ldots, 100$ and $t=151, \ldots, 200$.

We investigate the impact of increasing network dimensions on the estimation performance in terms of accuracy and consistency, by varying $N$ from 10 to 100 with an increment of 10 or one sub-block. The sample size $T$ is fixed to create the scenarios of dimensionality $N < T$ and $N \approx T$. The simulations were repeated 100 times. We computed factor-SVAR model-based estimates for the state sequence $\widehat{S}_t, t=1, \ldots, 200$ and the coefficient matrix for each state $\widehat{\bf \Phi}_{\bf y}^{[j]}$, using the estimation steps in Section 2. The number of factors was selected adaptively for each simulated data using BIC in \ref{Eq:BIC}. \\

\textit{\underline{Benchmark with K-means Clustering}}: We compare the performance of our factor-SVAR estimator with an recent approach based on K-means clustering of time-variant VAR coefficients proposed by \cite{Balqis2016}. Here, a sliding window is first employed to estimate the time-evolving directed connectivity, by fitting stationary VAR model to shifted short-time windows of fixed length to obtain time-dependent estimates of VAR coefficients matrices. We used a rectangular window with a bandwidth of 30 samples and shift of 1 sample. The relatively short-segments may render the traditional ordinary least-squares (LS) fits of large-dimensional VAR matrices inaccurate, due to insufficient information to estimate the huge number of parameters. Therefore, we used the $L_2$-regularized or ridge estimator which imposes a $L_2$ norm penalty on the AR coefficients in the LS regression, to obtain a better-conditioned estimate particularly in high-dimensional settings. The regularization parameter was set $\lambda=0.1$, as suggested by \cite{korobilis2013var} for VAR model estimation. Then, the K-means clustering algorithm is applied to the estimated time-variant VAR (TV-VAR) coefficients to partition the dynamic connectivity structure into the distinct states or regimes. As in \cite{Allen2012}, we used the L1 (Manhattan) distance which may be more effective for clustering high-dimensional data, compared to the L2 (Euclidean) distance. 

Our proposed SVAR approach has more advantages than the K-means clustering of time-variant VAR coefficients, as discussed in \cite{Balqis2016}. First, the sliding-window approach is limited by the choice of window size which is crucial: a large window leads to low statistical power for detecting abrupt and highly localized changes; a small window produces noisy estimates for smooth changes. In contrast, the SVAR model is capable of detecting changes at different time scales, both smooth and abrupt, avoiding the problems associated with fixed time windowing. Second, the K-means algorithm provides a `hard' assignment of time points into states and does not account for the temporal correlation structure. In contrast, the SVAR estimator generates `soft' state-time alignment by estimating sequentially, for each time point, the probability of the occupying states based on the entire observation time course. \\

\textit{\underline{Performance Measure}}: To measure the performance of the estimated VAR connectivity graphs within each regime, we computed for each simulation the total squared errors over all entries between the ground-truth and the estimators of the VAR coefficient matrix for each state $j = 1,2$, $\|\widehat{\bf \Phi}_{\bf y}^{[j]} - {\bf \Phi}_{\bf y}^{[j]}\|_F^2$, where $\|{\bf H}\|_F = {\tr({\bf H}'{\bf H})}^{1/2}$ denotes the Frobenius norm of matrix ${\bf H}$. To evaluate the connectivity regime change-point detection, we measure the percentage of correctly classified time points into the true states for each simulated time course. \\

\begin{figure}[!ht]
	\begin{minipage}[t]{1\linewidth}
		\centering
		\includegraphics[width=0.5\linewidth,keepaspectratio]{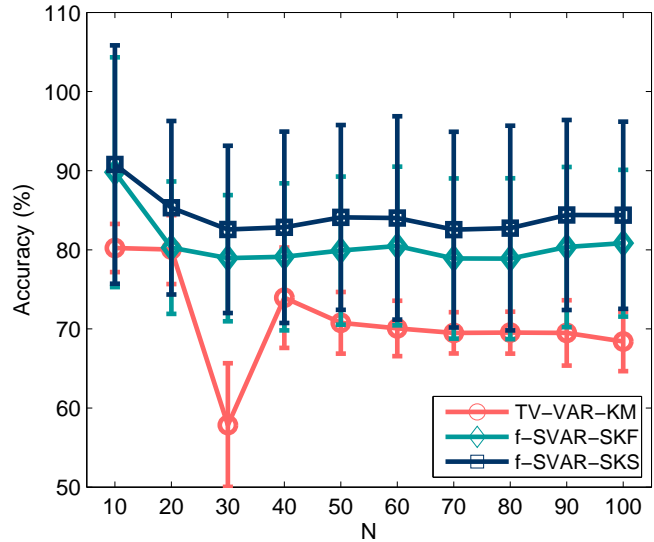}
	\end{minipage}
\caption{Accuracy of state classification of time-points obtained using K-means clustering, $\widehat{S}^{\text{KM}}_t$, switching KF, $\widehat{S}^{\text{SKF}}_t$ and switching KS, $\widehat{S}^{\text{SKS}}_t$, as a function of dimension $N$ for the simulated fMRI data from a regime-switching VAR(1) with $K=2$ states. Lines and error bars represent the averages and standard deviations over all replications.}
\label{Fig:State-acc}
\end{figure}

\textit{\underline{Results}}: Figure \ref{Fig:State-acc} plots the averages and standard deviations of the state classification accuracies for different estimators over all replications, as a function of dimension $N$. Both the factor-SVAR model-based estimates $\widehat{S}^{\text{SKF}}_t$ and $\widehat{S}^{\text{SKS}}_t$ perform better in regime segmentation than the K-means clustering, with substantially higher accuracy consistently for all $N$, albeit with higher standard deviations. The refined smoothed estimates $\widehat{S}^{\text{SKS}}_t$ based on the entire observations are more accurate than the filtered estimates $\widehat{S}^{\text{SKF}}_t$. Moreover, it can be seen that the accuracy of K-means clustering drops as $N$ increases, while for both the switching Kalman estimates, it tends to stabilize for high dimensions when $N \geq 30$. This may be because the regime partitioning was done based on the noisy estimates of high-dimensional TV-VAR coefficients fitted on short-windowed samples, compared to the lower-dimensional, reliably estimated subspace of factors in our approach. Another reason is the inherent limitation of the K-means algorithm itself neglecting temporal evolution of the connectivity states, which instead can be captured by the Markov chain of the switching model.

\begin{figure*}[!ht]
\centering
	\begin{minipage}[t]{0.5\linewidth}
		\centering
		\subfloat[]{\includegraphics[width=3.25in, height=2.25in,keepaspectratio]{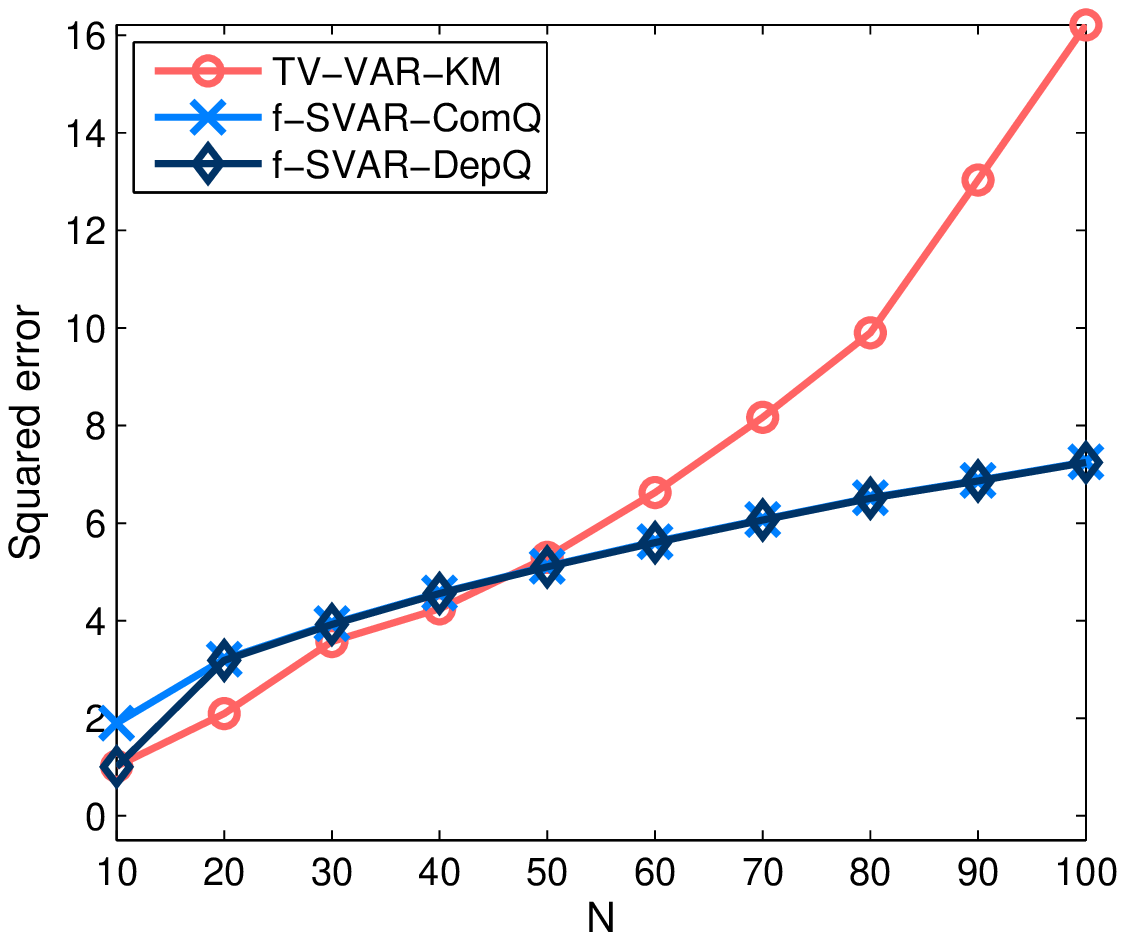}}
	\end{minipage}
	\hspace{-1cm}
	\begin{minipage}[t]{0.5\linewidth}
		\centering
		\subfloat[]{\includegraphics[width=3.25in, height=2.25in,keepaspectratio]{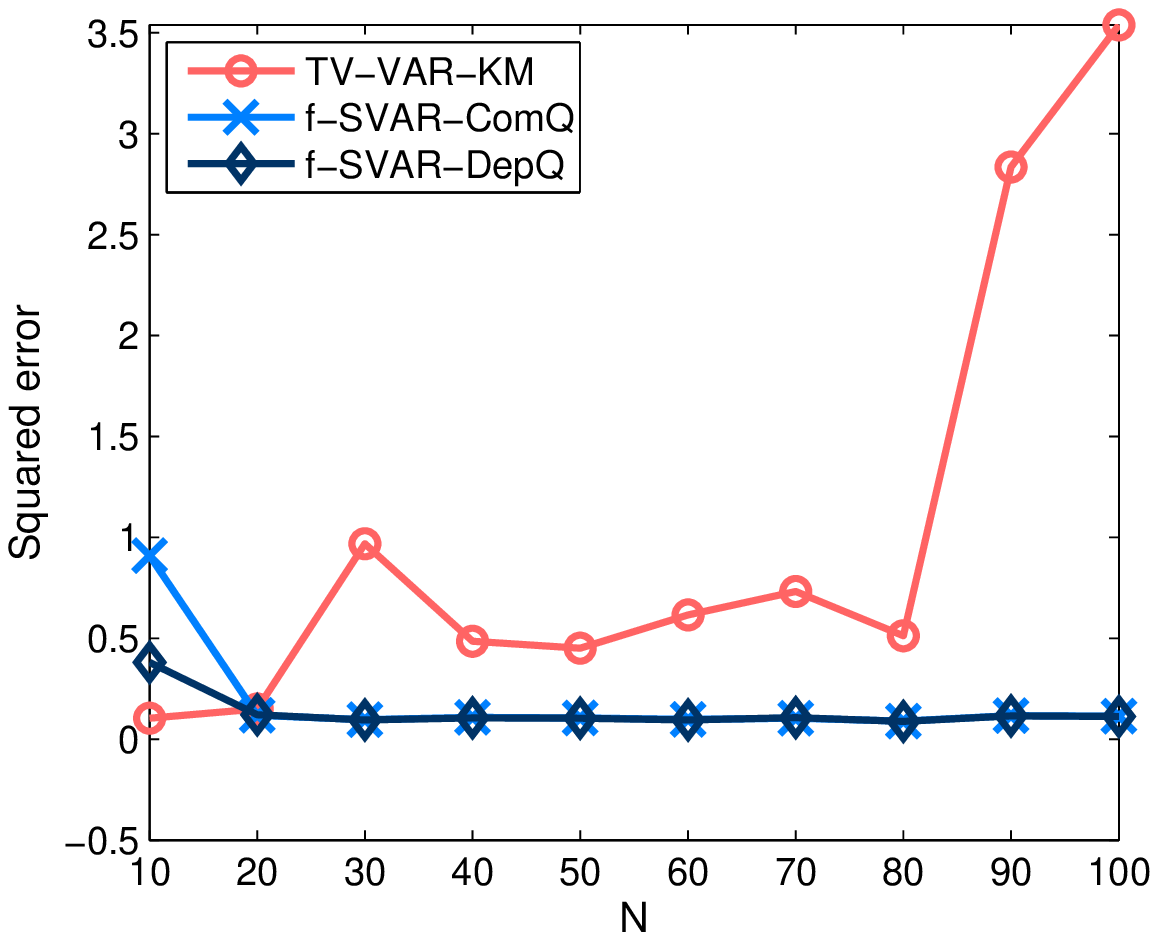}}
	\end{minipage}
\caption{(a) Averages and (b) standard deviations of squared estimation errors under Frobenius norm $\|\widehat{\bf \Phi}_{\bf y}^{[1]} - {\bf \Phi}_{\bf y}^{[1]}\|_F^2$ for the directed connectivity matrix at state $j=1$, using the K-means clustering with TV-VAR and the coupled and decoupled factor SVAR estimator, as a function of network dimension $N$.}
\label{Fig:SimError-A1}
\end{figure*}

\begin{figure*}[!ht]
\centering
	\begin{minipage}[t]{0.5\linewidth}
		\centering
		\subfloat[]{\includegraphics[width=3.25in, height=2.25in,keepaspectratio]{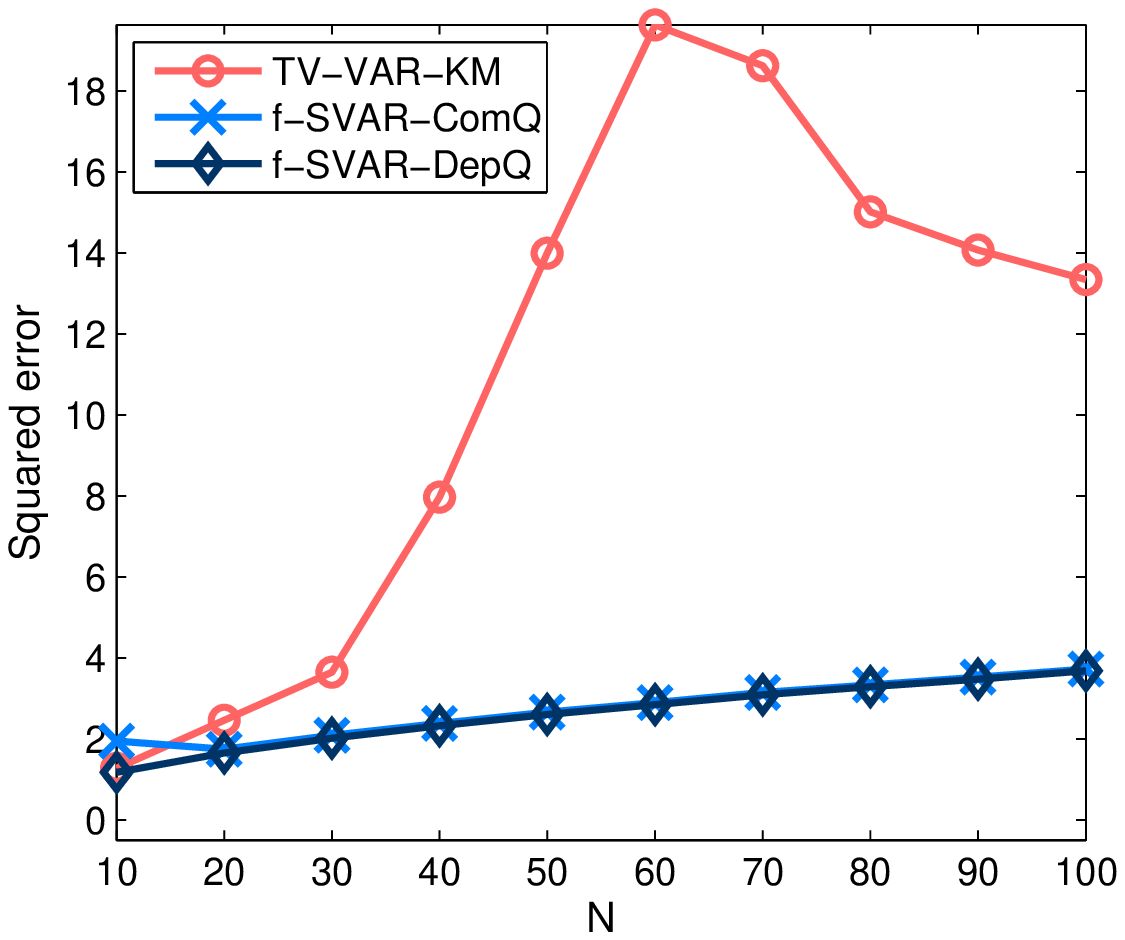}}
	\end{minipage}
	\hspace{-1cm}
	\begin{minipage}[t]{0.5\linewidth}
		\centering
		\subfloat[]{\includegraphics[width=3.25in, height=2.25in,keepaspectratio]{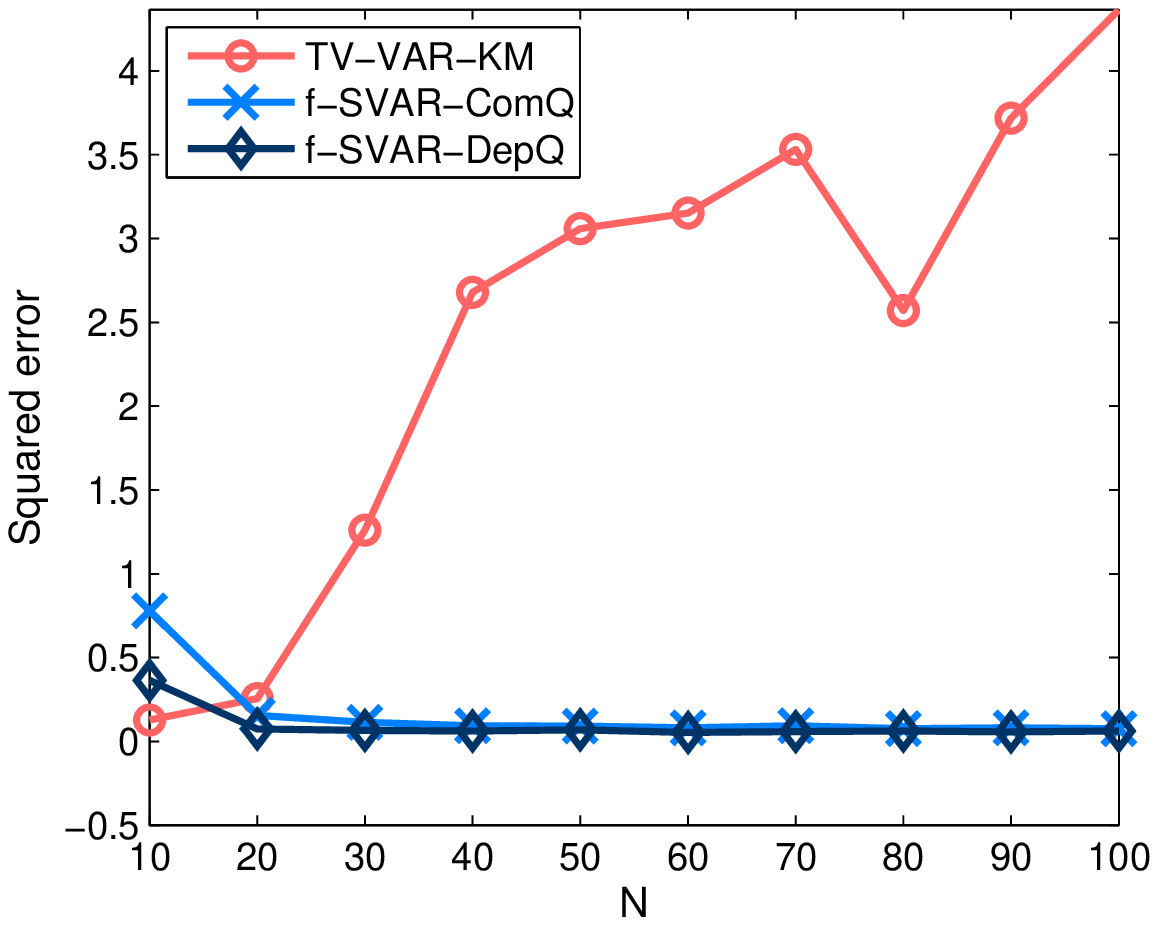}}
	\end{minipage}
\caption{(a) Averages and (b) standard deviations of squared estimation errors under Frobenius norm $\|\widehat{\bf \Phi}_{\bf y}^{[2]} - {\bf \Phi}_{\bf y}^{[2]}\|_F^2$ for the directed connectivity matrix at state $j=2$, using the K-means clustering with TV-VAR and the coupled and decoupled factor SVAR estimator, as a function of network dimension $N$.}
\label{Fig:SimError-A2}
\end{figure*}

Figure \ref{Fig:SimError-A1} and Figure \ref{Fig:SimError-A2} plots the estimation errors of the directed connectivity matrix for the two states by the K-means clustering-based and the factor-SVAR model-based procedures, for increasing network dimensions $N$. The results are averages and standard deviations over the 100 replications, which respectively indicate the accuracy (unbiasedness) and consistency of the estimator. It is shown that the f-SVAR subspace estimators clearly outperform the $L_2$-regularized VAR estimator based on K-means clustered regimes, for both states and particularly for large $N$, in terms of significantly lower estimation mean squared errors and standard errors, and only slightly underperformed when $N$ is small. We can also see a rapidly growing trend of estimation errors in the K-means-based $L_2$ estimator as $N$ increases and approaches the regime sample size. In contrast, the robustness of the proposed f-SVAR estimators in high-dimensional settings is evident from the slower error rates (Figure \ref{Fig:SimError-A1}-\ref{Fig:SimError-A1}(a)) and the constancy of standard errors over the increased dimensions (Figure \ref{Fig:SimError-A1}-\ref{Fig:SimError-A1}(b)). These results can be explained by the more accurate regime segmentation by the SKS conditioned on the EM-estimated parameters as shown in \ref{Fig:State-acc}, and improved consistency of the factor-based estimator over the ridge estimator for high-dimensional VAR coefficient matrix in each regime. The asymptotic theory of our proposed estimator such as convergence rates will be further studied in future work. Among the f-SVAR methods, both the coupled (common ${\bf Q}$) and decoupled (regime-dependent ${\bf Q}$) estimators perform comparably, despite slight superiority of the later. This suggests that the difference in directed connectivity structure based on a block-diagonal VAR model is mostly explained by inter-dependence in the factors, and less so in the projection of the underlying subspace. Hence, it can be sufficiently approximated by regime-dependent factor process, with a constant factor loading matrix across regimes.

\clearpage

\section{Application to Estimating Dynamic Brain Connectivity}

In this section, we shall apply the proposed f-SVAR approach to estimating time-evolving effective connectivity in high-dimensional resting-state fMRI data, characterized by abrupt transition of underlying quasi-stable brain states.

\subsection{Resting-state fMRI Data}

\textit{1) Data acquisition:}We studied the resting-state fMRI data of 10 subjects from the first scan of a dataset publicly available at NITRC (http://www.nitrc.org/projects/trt). A Siemens Allegra 3.0-Tesla scanner was used to obtain three resting-state scans for each subject. During scans, the subjects were asked to relax and keep their eyes open. BOLD functional images were acquired using a T2-weighted gradient-echo planar imaging (EPI) sequence (TR = 2000 ms; time echo (TE) = 25 ms; flip angle (FA) = $90^{\circ}$; field of view (FOV) = 192 mm; voxel size = $3\times3\times3$ mm$^3$; matrix $64\times64$; number of slices = 39). A time-series of $T = 197$ EPI volumes was collected for each scan.

\textit{2) Preprocessing:} The data were preprocessed using the AFNI and FSL software packages as in \cite{Fiecas2013}. The steps included (1) Motion correction using six-parameter rigid body transformation, normalized correlation as cost-function and referencing to the middle volume; (2) Spatial normalization to the Montreal Neurological Institute (MNI) template; (3) Probabilistic segmentation of the brain to obtain white matter and cerebrospinal fluid (CSF) probabilistic maps, thresholded at 0.99. (4) Removal of the nuisance signals, namely the six motion parameters, white matter and CSF signals, and the global signal. (5) Spatial smoothing with a 6 mm full-width half-maximum (FWHM) Gaussian kernel.

\textit{3) Parcellation:} We used the automated anatomical labeling (AAL) atlas to obtain an anatomical parcellation of the whole brain into 90 ROIs with 45 regions in each hemisphere . In this study, the ROIs were grouped into six pre-defined resting-state system networks (RSNs) of similar anatomical and functional properties, based on the templates in \cite{Allen2012,Li2011}. The considered RSNs include sub-cortical (SCN), AN: auditory (AN), sensorimotor (SMN), visual (VN), attentional (ATN) and default mode network (DMN). We followed the ROI abbreviations in \cite{Salvador2005}.

\subsection{Results}

We analyzed the dynamic states of large-scale effective brain connectivity in the resting state. We fitted a three-state factor-SVAR(1) model using the EM algorithm to the resting-state fMRI time series concatenated for all subjects, to identify the state transitions and the high-dimensional directed dependencies within each state which are assumed to be shared across subjects (as measured respectively by the SKS-estimated state-time sequence and state-dependent VAR coefficient matrices). Here, the decoupled SVAR subspace estimator was used, and the number of factors selected for this data by using BIC was $\hat{r}=14$. We used the VAR model order of one, as typically assumed for fMRI data \cite{Valdes-Sosa2005}.

Figure \ref{Fig:state-net-mat} shows the estimated whole-brain directed connectivity matrices between ROIs for three distinct states, and the corresponding within-network connectivity graphs for three selected RSNs. Only significant connections are shown, tested based on the asymptotic normality of the factor-VAR coefficient estimator in (\ref{Eq:Asym-fVAR}), at $\alpha = 0.05$ with Bonferroni correction. Our method identifies the modular organization of the resting-state networks over all states, where ROIs within a functionally relevant network tend to be densely connected, but sparsely connected between different networks, particularly pronounced in VN, DMN and SMN. This characteristic has been reported in previous studies of static fMRI functional connectivity, e.g. \cite{Ferrarini2009}. In consistency with findings in dynamic functional connectivity states \cite{Allen2012,Hutchison2013}, our results also show the distinct large-scale connectivity patterns across different brain states in terms of variability in both the strength and sign of the connectivity and the network modularity. More interestingly, our method further reveals state-related difference in the directionality of the connections not reported previously, as evident from the asymmetry of the estimated VAR coefficient matrices. We discuss few apparent patterns that differ between the effective connectivity states. It is shown that the states are differentiated by the ROI-wise connectivity for both between-networks and within-networks. For the within-network connectivity, we observe the strongest connections between ROIs in state 1 for all the three RSNs, generally. For the sensorimotor networks, the directed interactions between central regions in the primary motor cortex is the strongest in state 1, which however shows disrupted connections with the parietal regions. In state 2, we found strong uni-directional influences from both the superior parietal nodes (SPG.L and SPG.R) to the supplementary motor area (SMA) with negative correlation (as indicated by blue edges), which are not present in states 1 and 3. 
For the attentional networks, we identified the lateral frontal-parietal network (similar to ventral attention network \cite{Vincent2008}) between regions e.g. middle frontal gyrus and inferior parietal gyrus in all states, with the strongest connections occurred in state 1. However, we found denser directed information flows across both left and right hemispheres in states 2 and 3, compared to state 1. Particularly, it is interesting that the cross-hemisphere connections between parietal regions detected in states 2 and 3 were completely absent in state 1. For the default mode networks, state 1 also reveal the strongest and densest connections between ROIs related to posterior cingulate cortex (PCC)/precuneus, medial prefrontal cortex and the left and right inferior parietal lobule, with PCC correctly identified as a major hub of the DMN, strongly connected with other regions, as reported in numerous studies \cite{FranssonMarrelec2008}.

\begin{figure*}[!ht]
\centering
\captionsetup[subfigure]{farskip=0pt}
\setcounter{subfigure}{0}
\captionsetup[subfigure]{labelformat=empty,position=top,textfont=bf,font=small}

\begin{tabular}{c@{}cccc@{}}
\raisebox{-1.3cm}{\subfloat{\rotatebox{90}{\small\textbf{State 1}}}} \hspace{0.1cm} &
\subfloat[Connectivity Matrix]{\includegraphics[width=0.285\linewidth,keepaspectratio]{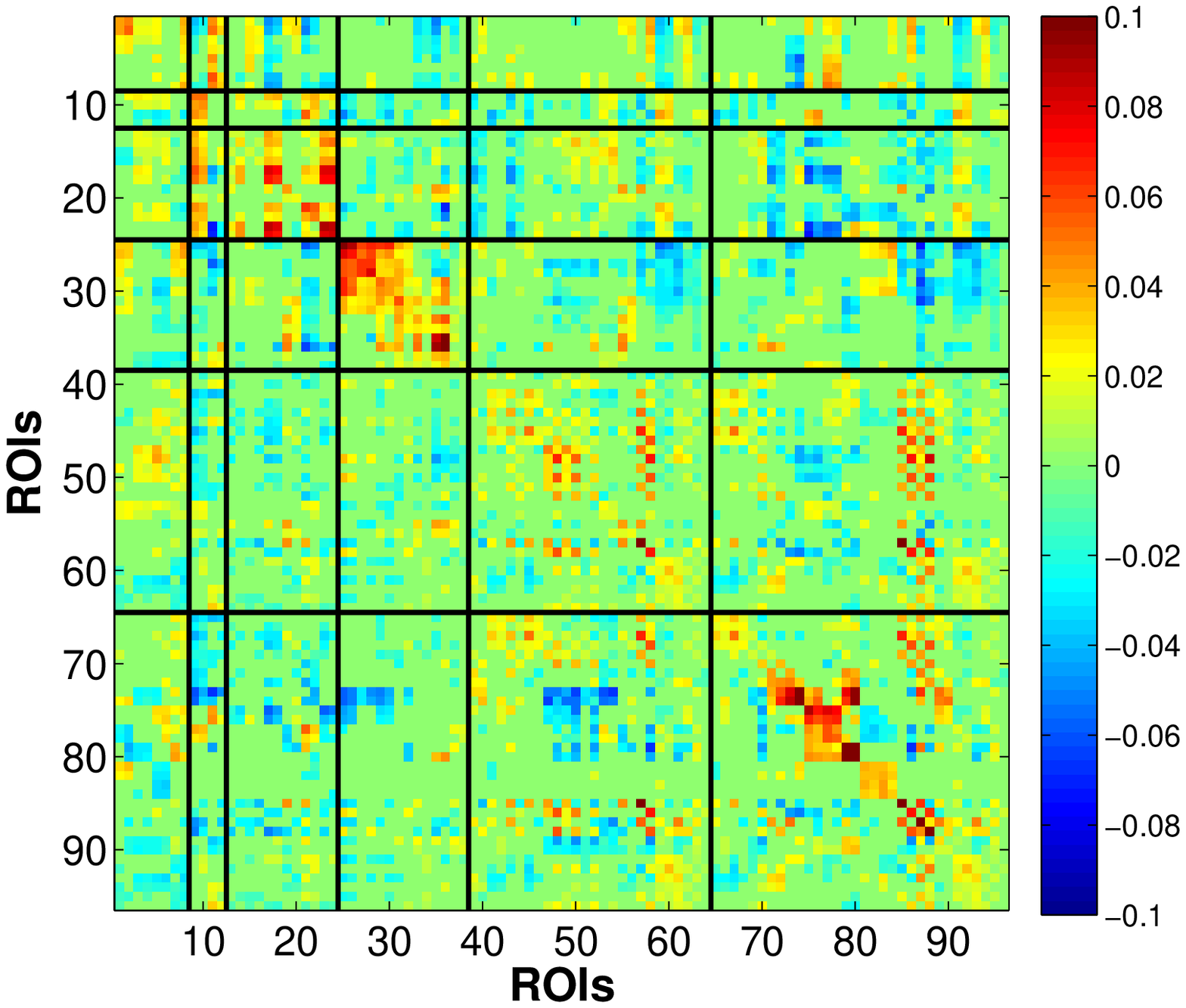}} \hspace{-0.2cm} &
\subfloat[Sensorimotor]{\includegraphics[width=0.21\linewidth,keepaspectratio]{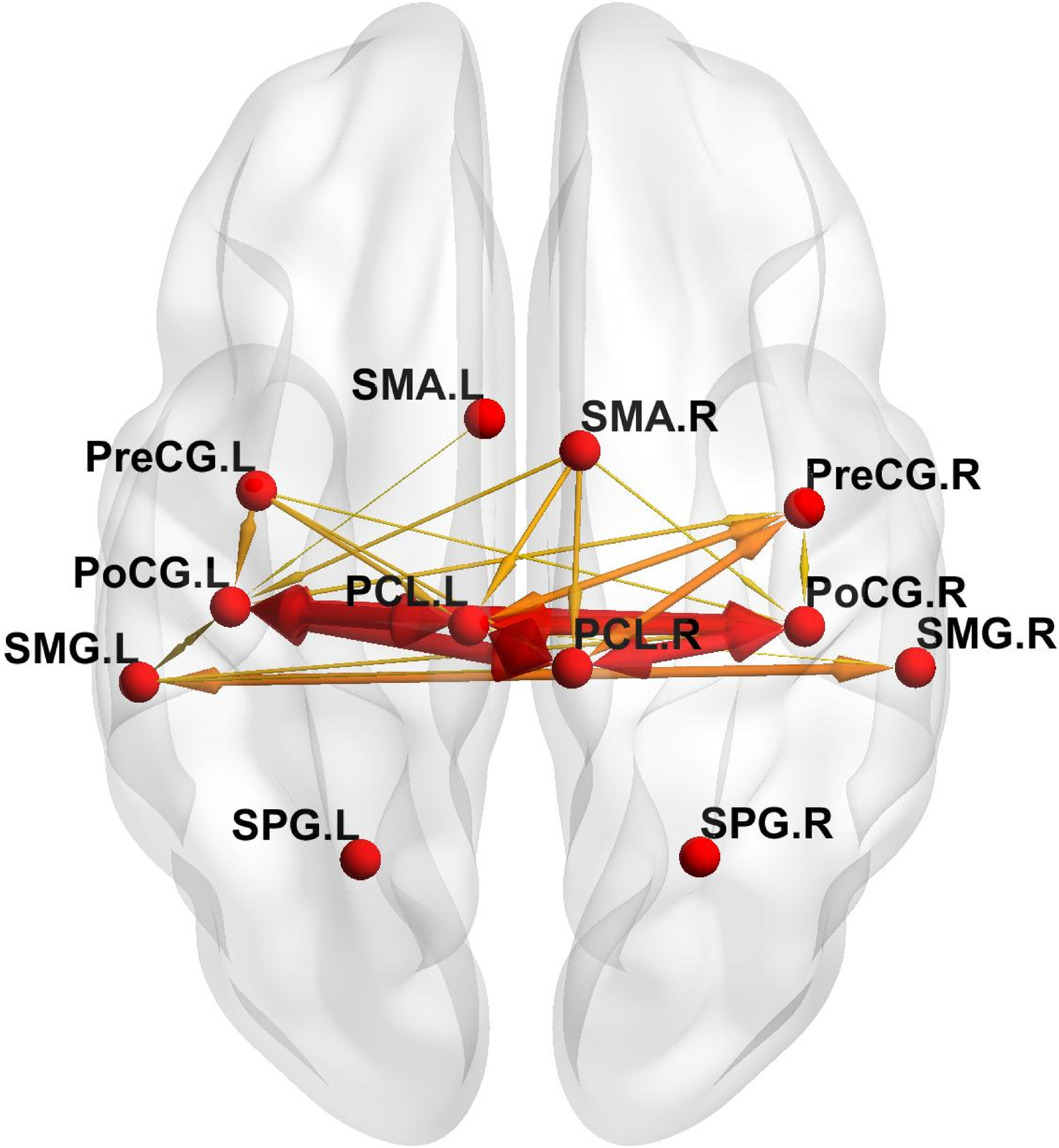}} \hspace{-0.5cm} &
\subfloat[Attentional]{\includegraphics[width=0.21\linewidth,keepaspectratio]{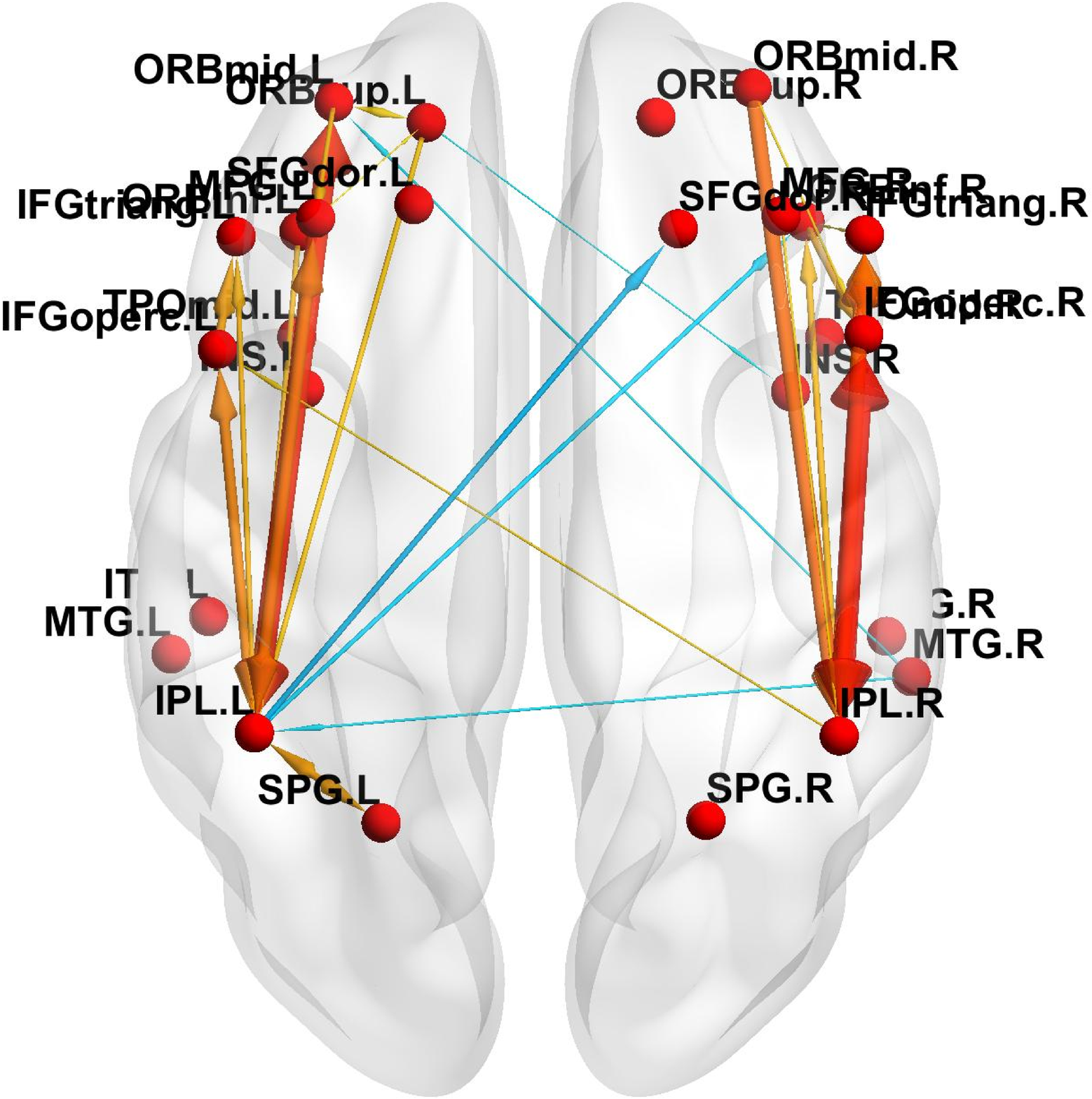}} \hspace{-0.5cm} &
\subfloat[Default Mode]{\includegraphics[width=0.21\linewidth,keepaspectratio]{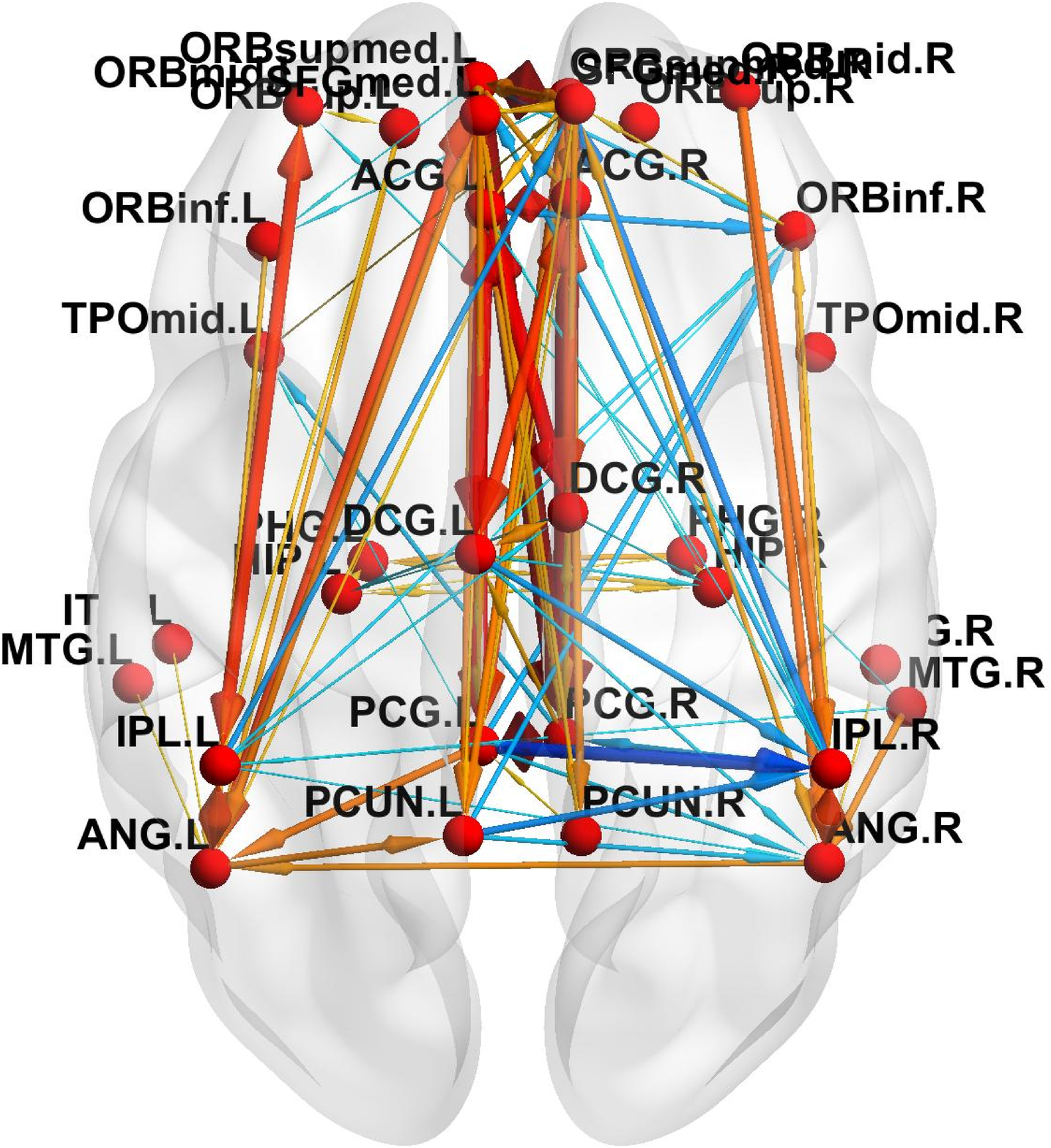}}  \vspace{-0.4cm} \\

\raisebox{-1.3cm}{\subfloat{\rotatebox{90}{\small\textbf{State 2}}}} \hspace{0.1cm} &
\subfloat[]{\includegraphics[width=0.285\linewidth,keepaspectratio]{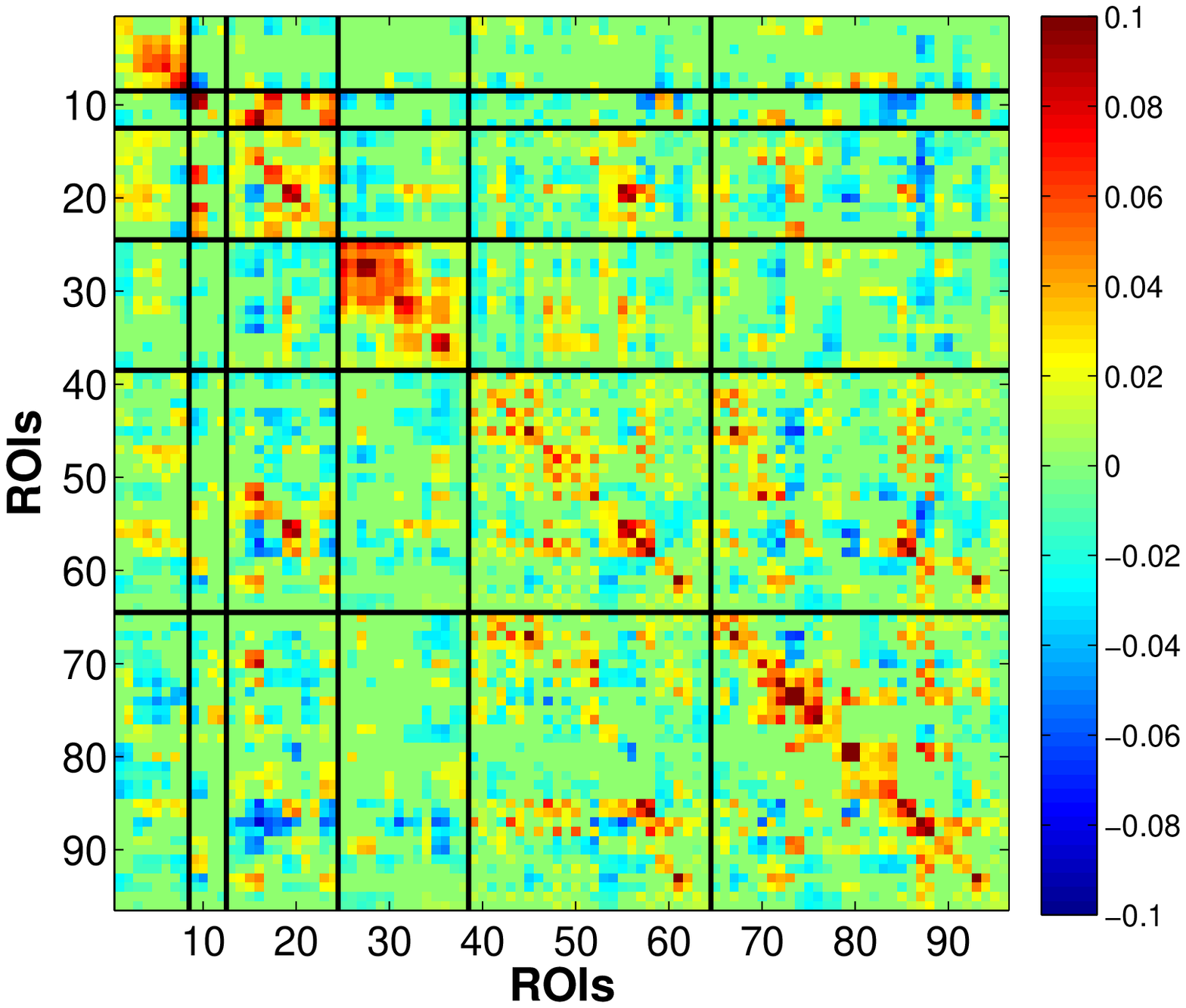}} \hspace{-0.2cm} &
\subfloat[]{\includegraphics[width=0.21\linewidth,keepaspectratio]{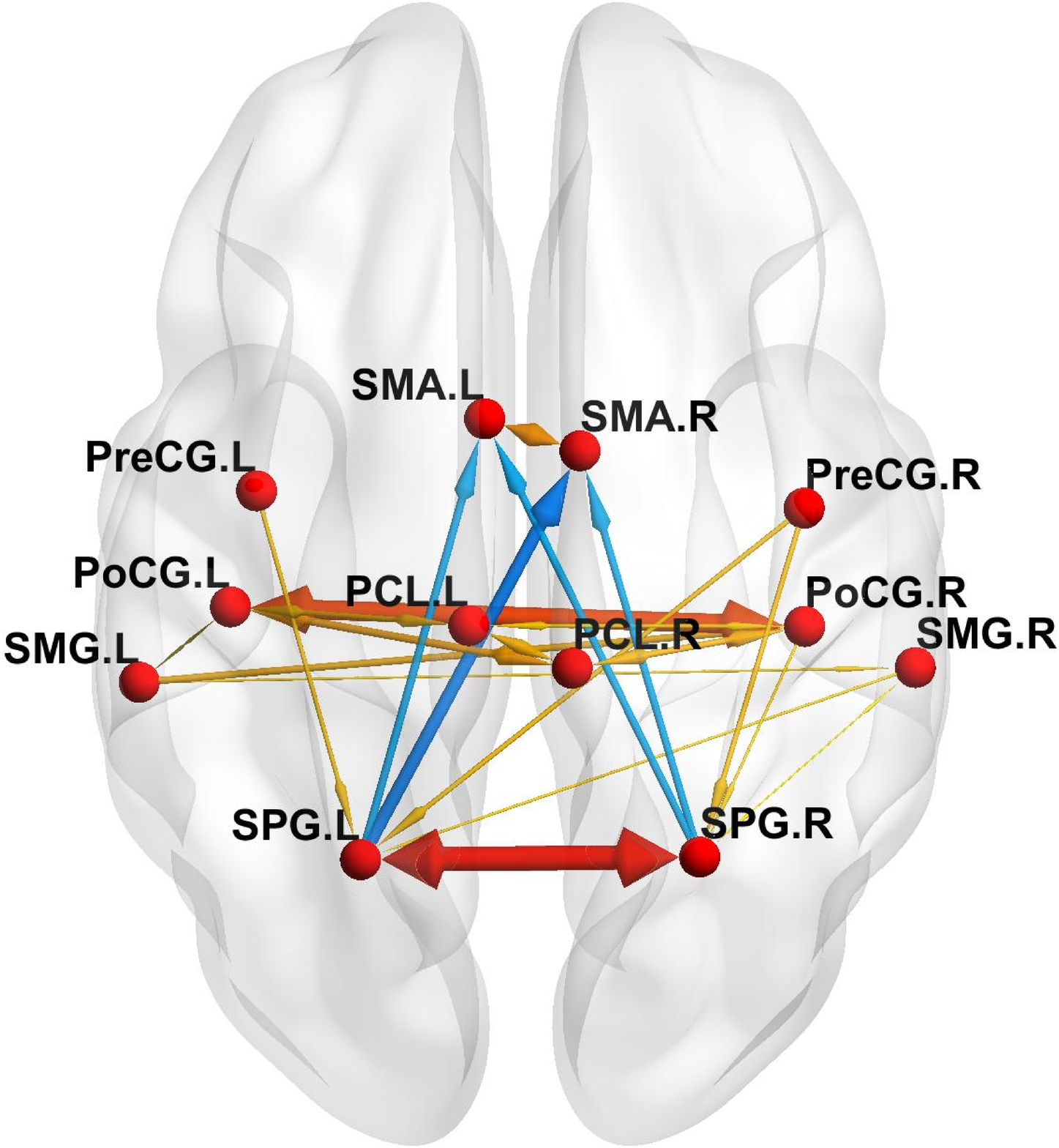}} \hspace{-0.5cm} &
\subfloat[]{\includegraphics[width=0.21\linewidth,keepaspectratio]{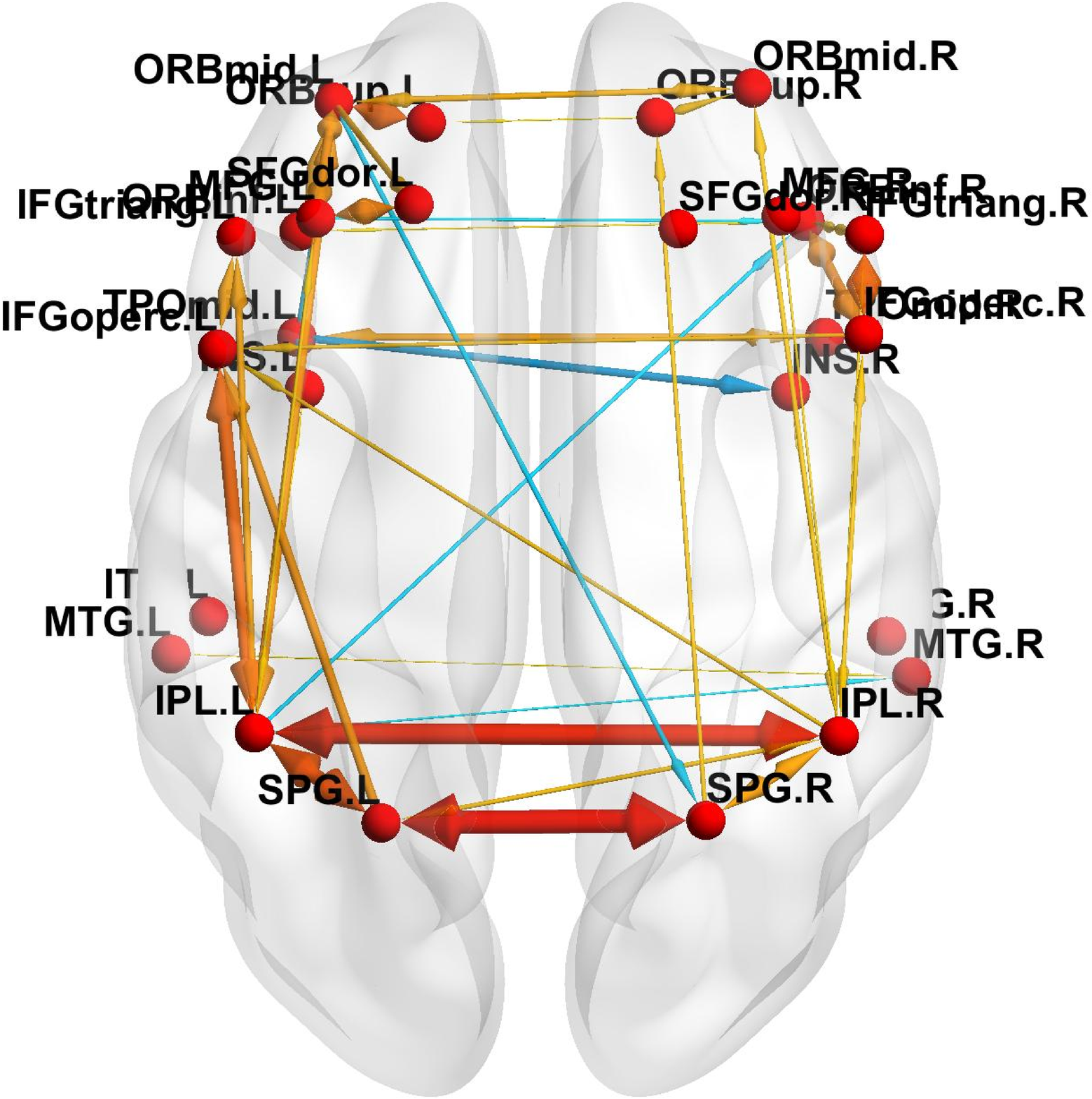}} \hspace{-0.5cm} &
\subfloat[]{\includegraphics[width=0.21\linewidth,keepaspectratio]{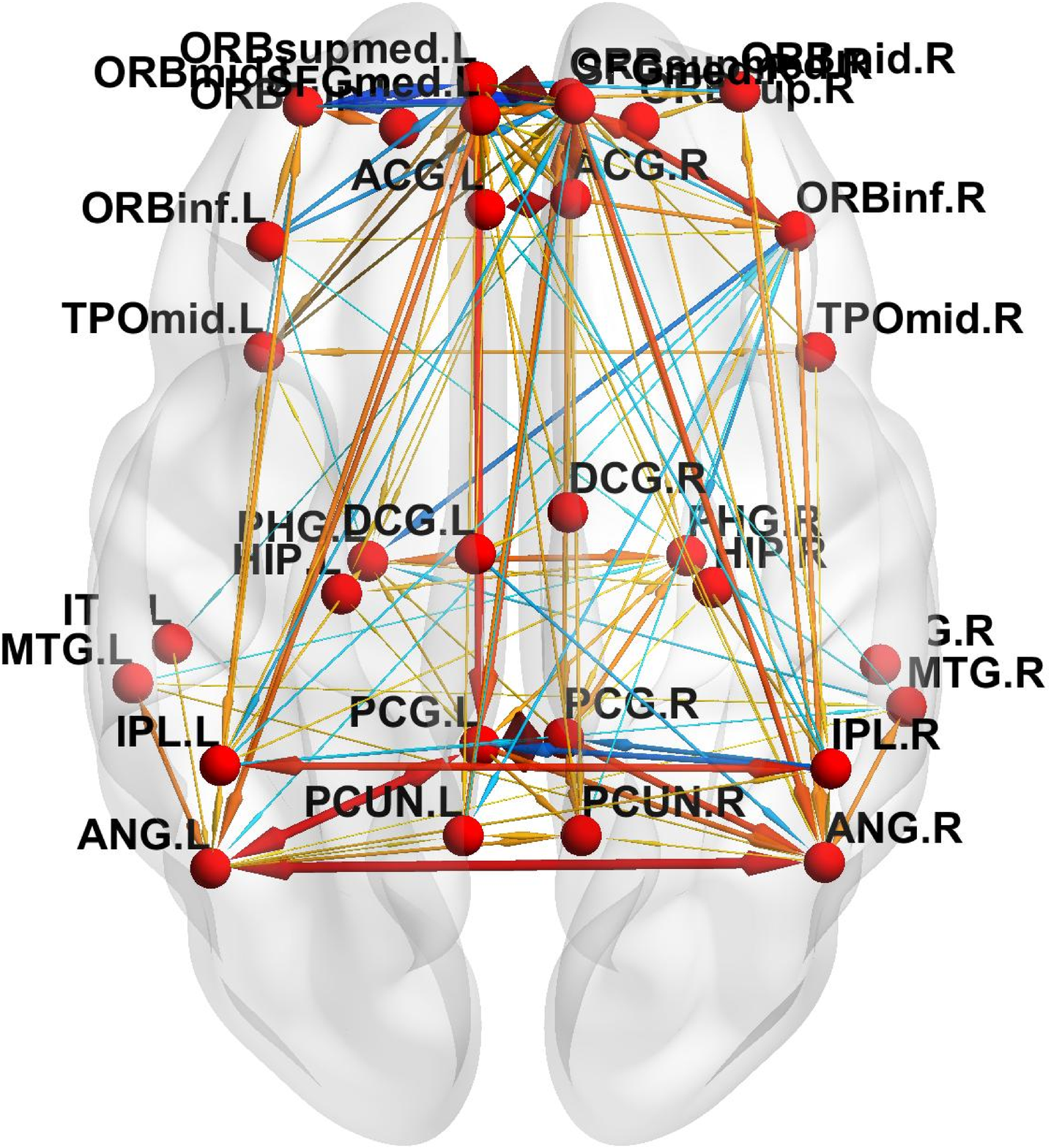}}  \vspace{-0.4cm} \\

\raisebox{-1.3cm}{\subfloat{\rotatebox{90}{\small\textbf{State 3}}}} \hspace{0.1cm} &
\subfloat[]{\includegraphics[width=0.285\linewidth,keepaspectratio]{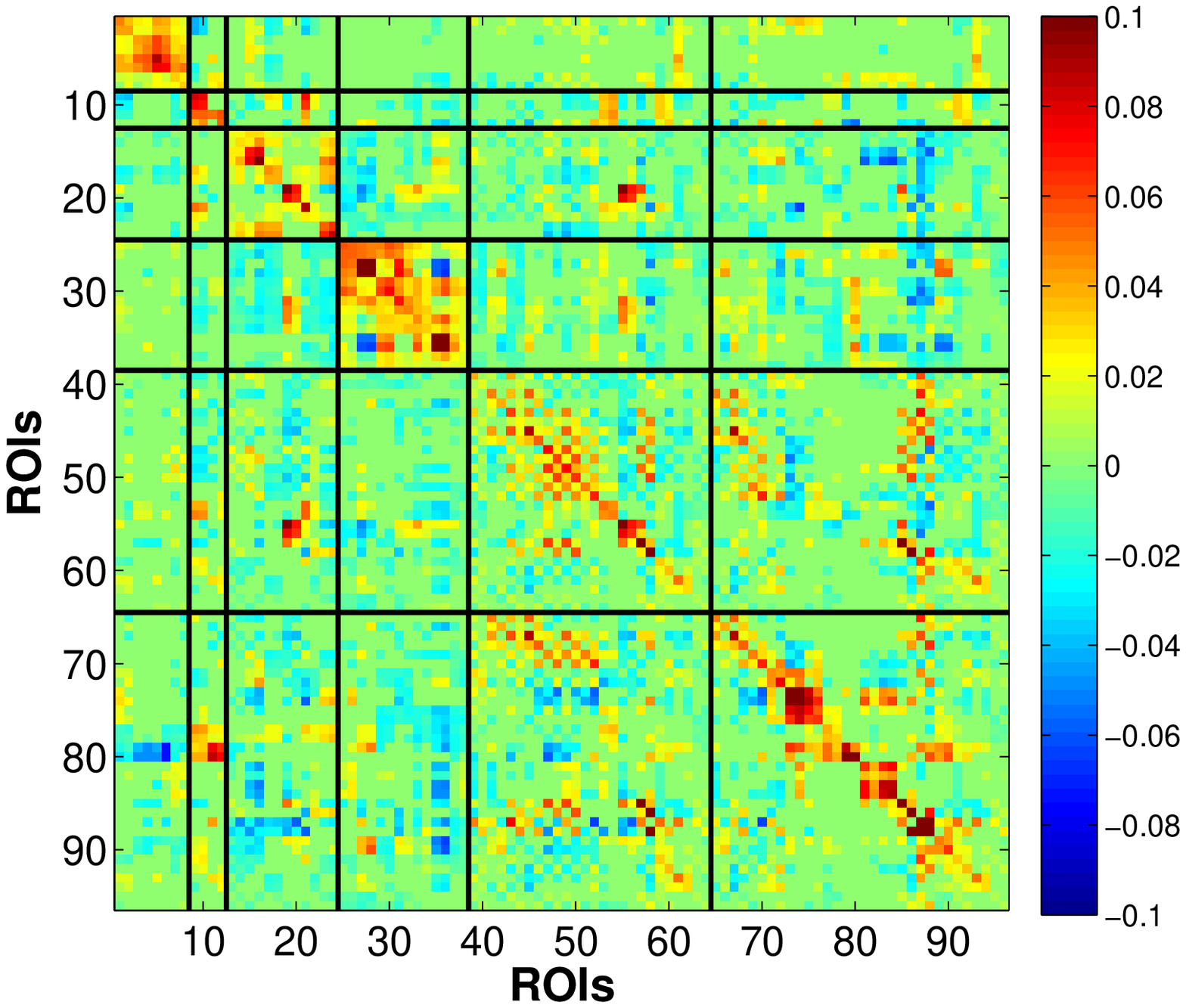}} \hspace{-0.2cm} &
\subfloat[]{\includegraphics[width=0.21\linewidth,keepaspectratio]{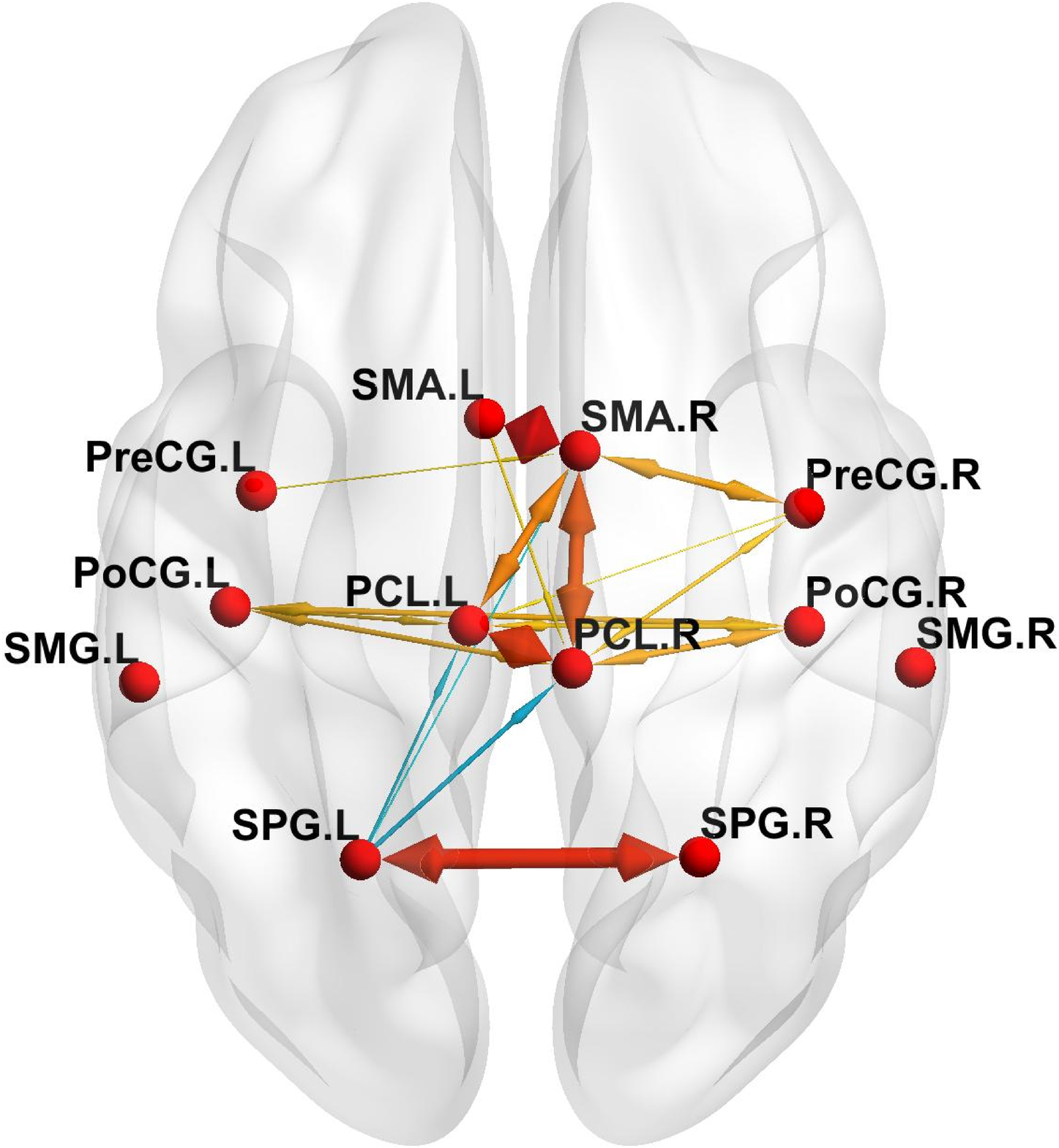}} \hspace{-0.5cm} &
\subfloat[]{\includegraphics[width=0.21\linewidth,keepaspectratio]{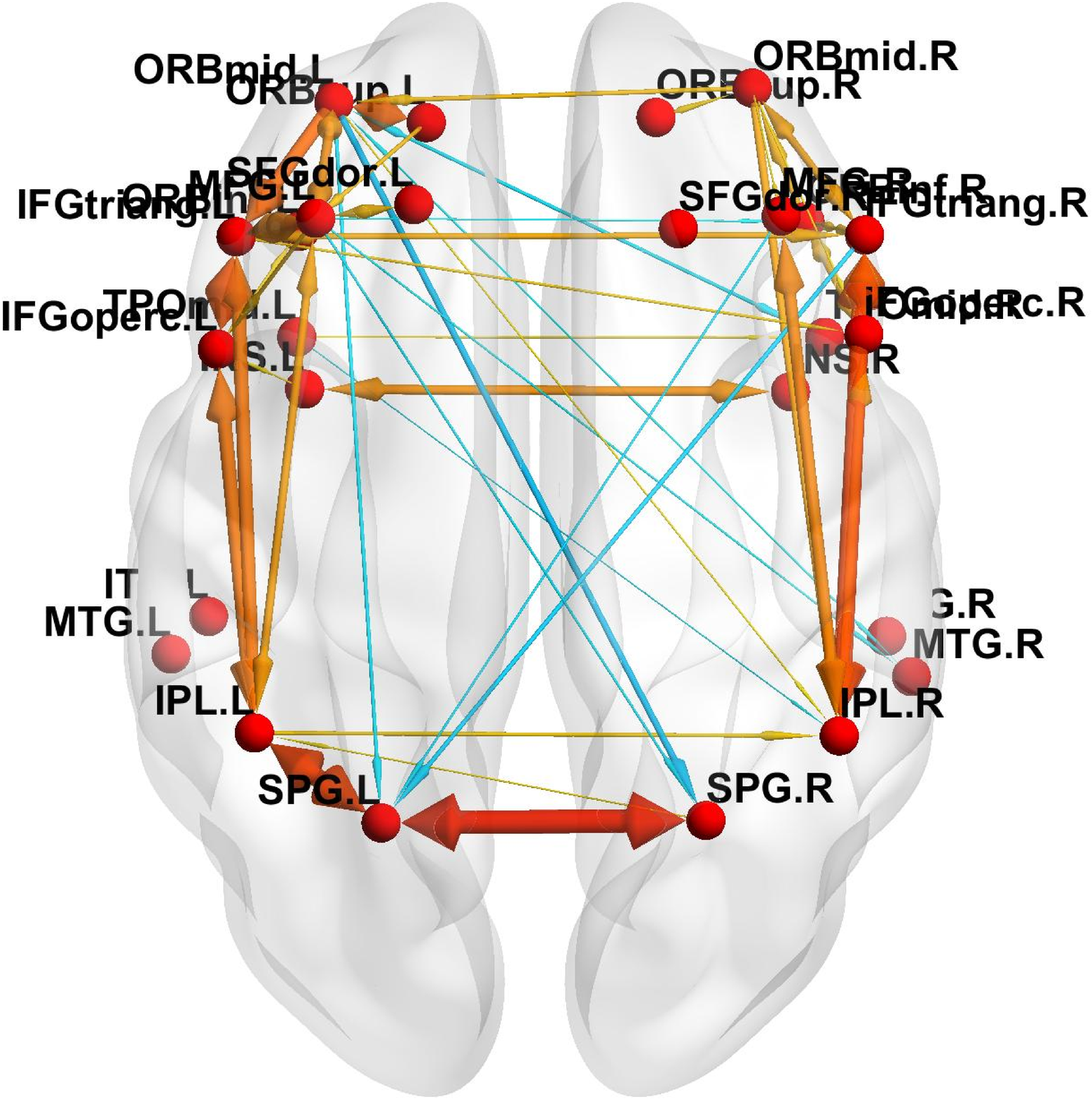}} \hspace{-0.5cm} &
\subfloat[]{\includegraphics[width=0.21\linewidth,keepaspectratio]{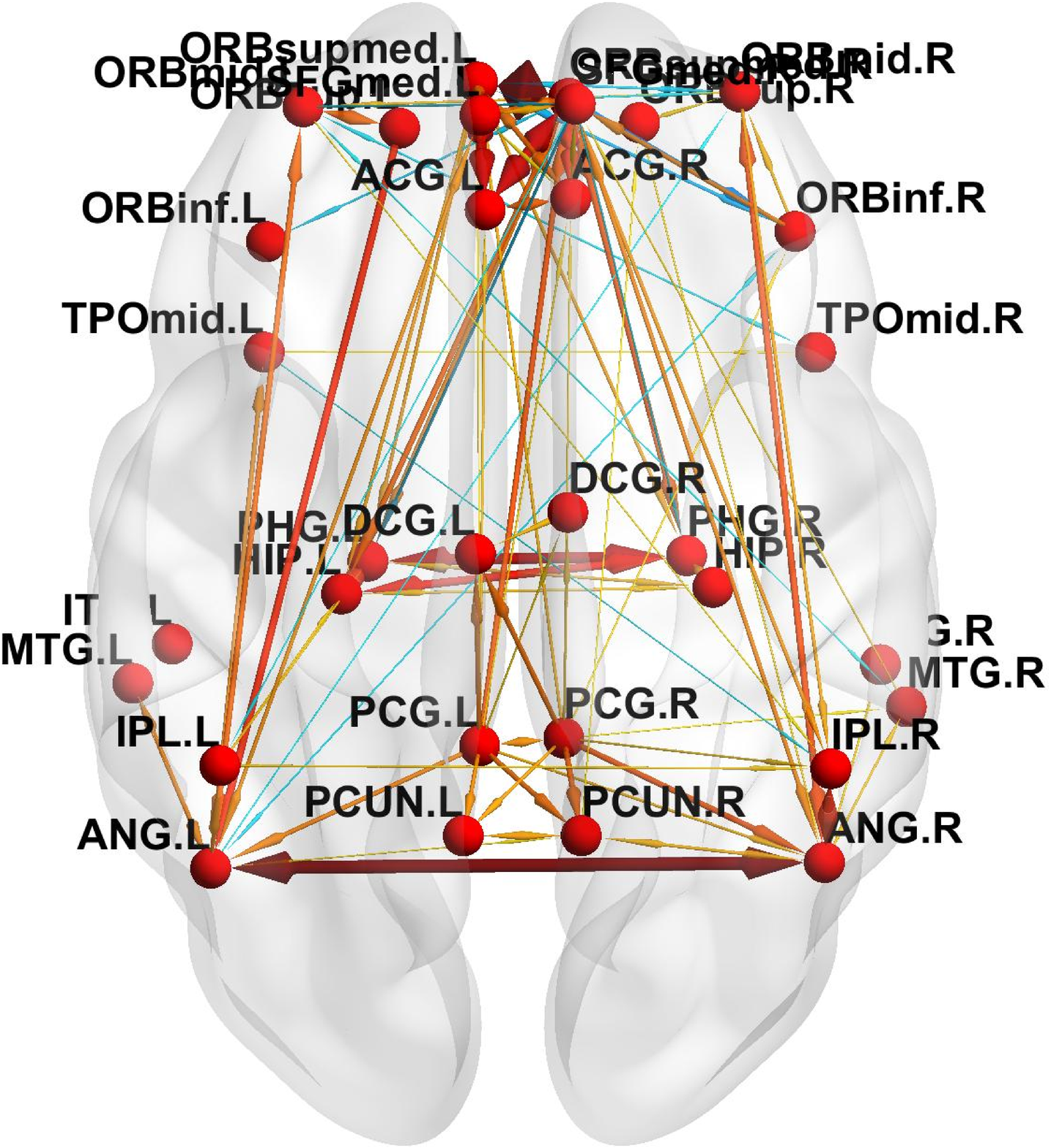}} \\
\end{tabular}
\caption{Effective connectivity states in the resting-state fMRI data across 10 subjects identified by the factor SVAR approach, show distinct large-scale connectivity patterns across three states: VAR coefficient matrix representations of the whole-brain connectivity between ROIs and the topological representations of within-network connectivity for three resting-state networks (RSNs). The 90 brain ROIs are grouped with overlapping into six RSNs: sub-cortical (SCN), auditory (AN), sensorimotor (SMN), visual (VN), attentional (ATN) and default mode network (DMN). The partitions are indicated by lines. The AR coefficient entries shown are significantly different from zero at level $\alpha = 0.05$ with Bonferroni correction for multiple testings. Edges represents strong connections with absolute AR coefficient than a threshold of 0.03, and arrows indicate the directionality of the connections.}

\end{figure*}\label{Fig:state-net-mat}

\begin{figure}[!ht]
\captionsetup[subfigure]{labelformat=empty}
	\begin{minipage}[t]{1\linewidth}
		\centering
		\subfigure[\label{Fig:st_sub_1}]{\includegraphics[width=0.5\linewidth,keepaspectratio]{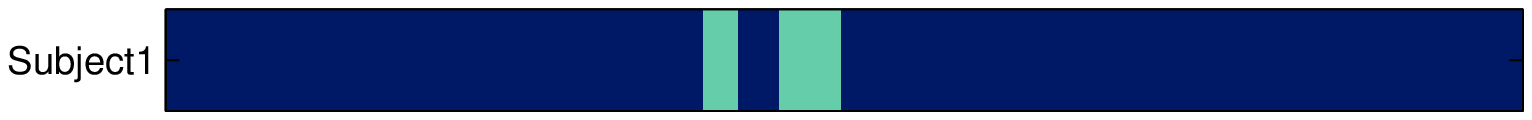}}
	\vspace{-0.63cm}
	\end{minipage}
	
	\begin{minipage}[t]{1\linewidth}
		\centering
		\subfigure[\label{Fig:st_sub_2}]{\includegraphics[width=0.5\linewidth,keepaspectratio]{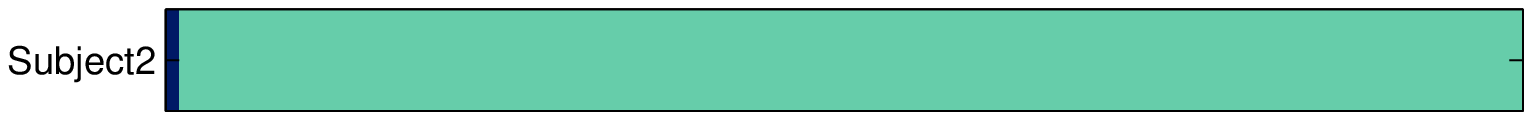}}
		\vspace{-0.63cm}
	\end{minipage}
	
	\begin{minipage}[t]{1\linewidth}
		\centering
		\subfigure[\label{Fig:st_sub_3}]{\includegraphics[width=0.5\linewidth,keepaspectratio]{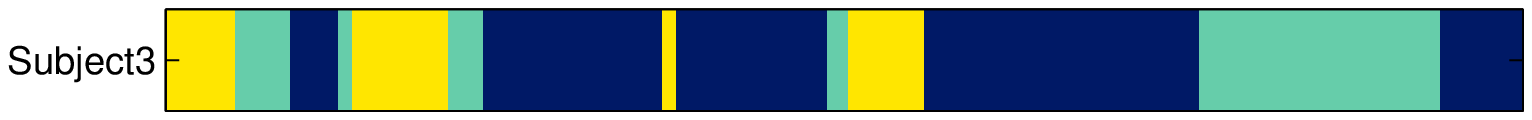}}
		\vspace{-0.63cm}
	\end{minipage}
	
	\begin{minipage}[t]{1\linewidth}
		\centering
		\subfigure[\label{Fig:st_sub_4}]{\includegraphics[width=0.5\linewidth,keepaspectratio]{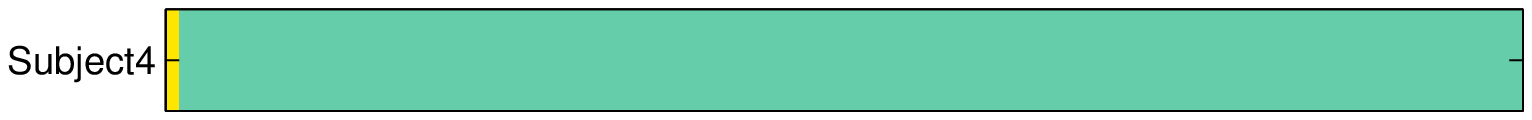}}
		\vspace{-0.63cm}
	\end{minipage}

	\begin{minipage}[t]{1\linewidth}
		\centering
		\subfigure[\label{Fig:st_sub_5}]{\includegraphics[width=0.5\linewidth,keepaspectratio]{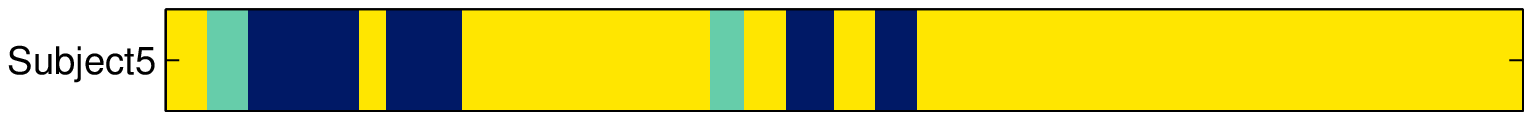}}
		\vspace{-0.63cm}
	\end{minipage}
	
	\begin{minipage}[t]{1\linewidth}
		\centering
		\subfigure[\label{Fig:st_sub_6}]{\includegraphics[width=0.5\linewidth,keepaspectratio]{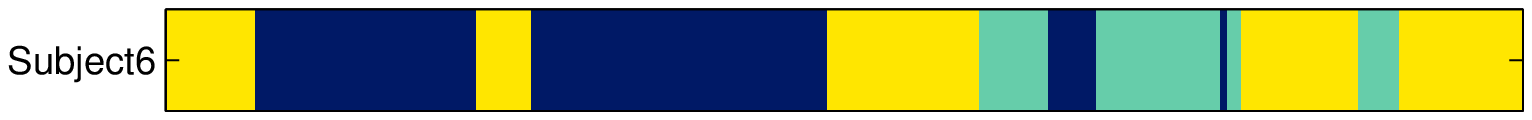}}
		\vspace{-0.63cm}
	\end{minipage}
	
	\begin{minipage}[t]{1\linewidth}
		\centering
		\subfigure[\label{Fig:st_sub_7}]{\includegraphics[width=0.5\linewidth,keepaspectratio]{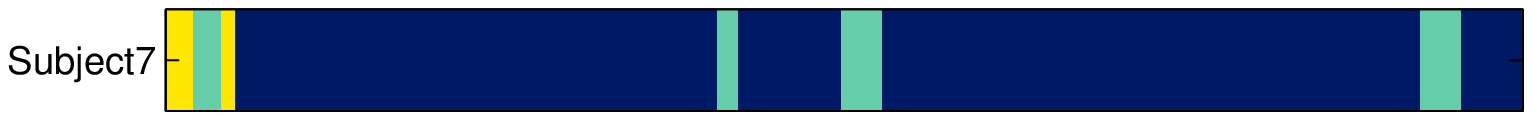}}
		\vspace{-0.63cm}
	\end{minipage}
	
	\begin{minipage}[t]{1\linewidth}
		\centering
		\subfigure[\label{Fig:st_sub_8}]{\includegraphics[width=0.5\linewidth,keepaspectratio]{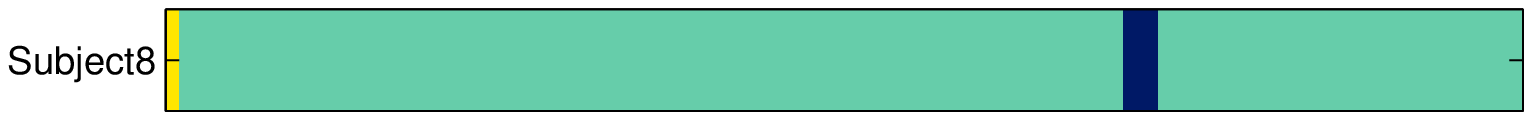}}
		\vspace{-0.63cm}
	\end{minipage}
	
	\begin{minipage}[t]{1\linewidth}
		\centering
		\subfigure[\label{Fig:st_sub_9}]{\includegraphics[width=0.5\linewidth,keepaspectratio]{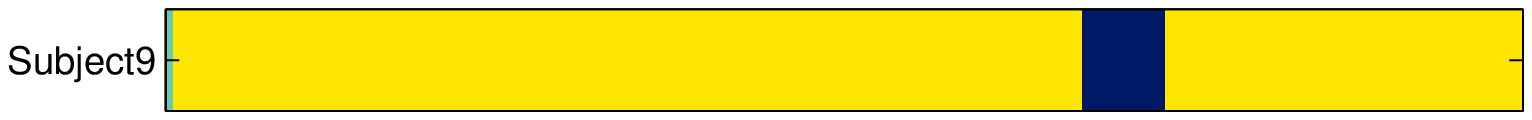}}
		\vspace{-0.63cm}
	\end{minipage}
	
	\begin{minipage}[t]{1\linewidth}
	\hspace{-0.23cm}
		\centering
		\subfigure[\label{Fig:st_sub_10}]{\includegraphics[width=0.5025\linewidth,keepaspectratio]{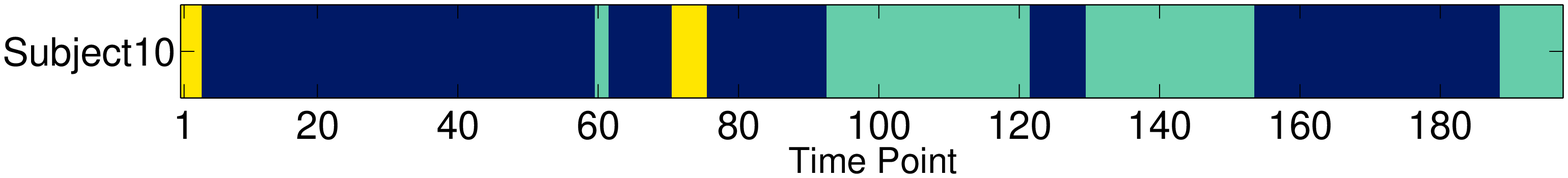}}
	\end{minipage}
	\vspace{-0.2in}
\caption{Tracking the temporal changes in effective connectivity states in fMRI data across 10 subjects during the resting state. State 1: yellow, State 2: blue, State 3: green.}
\label{Fig:states_subs}

\end{figure}

\begin{figure*}[!ht]
\captionsetup[subfigure]{labelformat=empty}
	\begin{minipage}[t]{1\linewidth}
		\centering
		\subfigure[(a)\label{Fig:fMRI-signals}]{\includegraphics[width=0.5\linewidth,keepaspectratio]{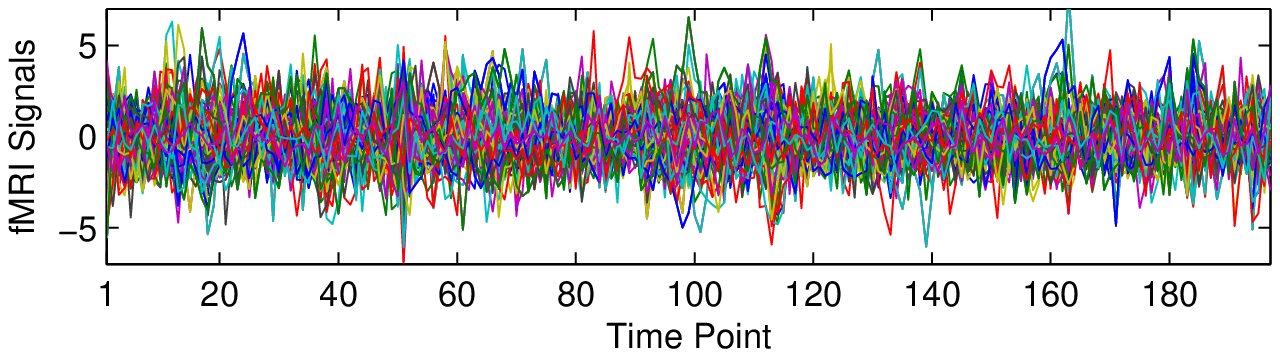}}
	\end{minipage}
	\begin{minipage}[t]{1\linewidth}
		\centering
		\subfigure[(a)\label{Fig:fMRI-factor-signals}]{\includegraphics[width=0.5\linewidth,keepaspectratio]{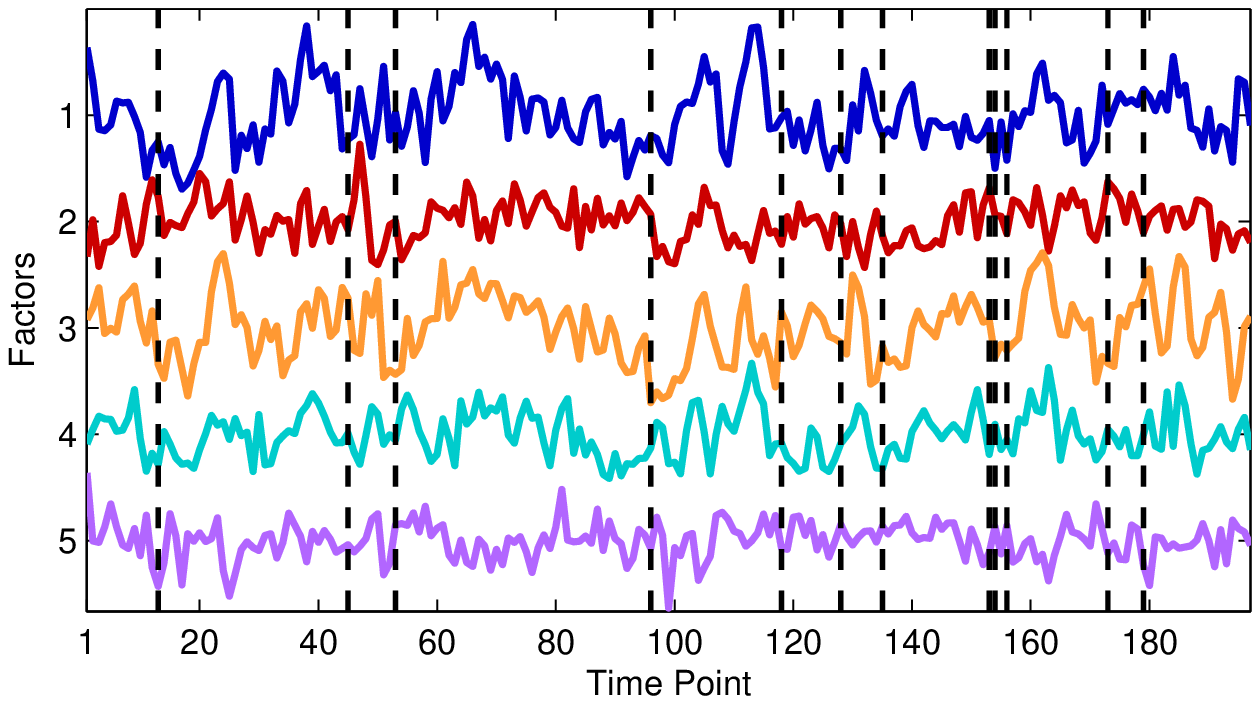}}
	\end{minipage}
	\begin{minipage}[t]{1\linewidth}
		\centering
		\subfigure[\label{Fig:SKF-states}]{\includegraphics[width=0.5\linewidth,keepaspectratio]{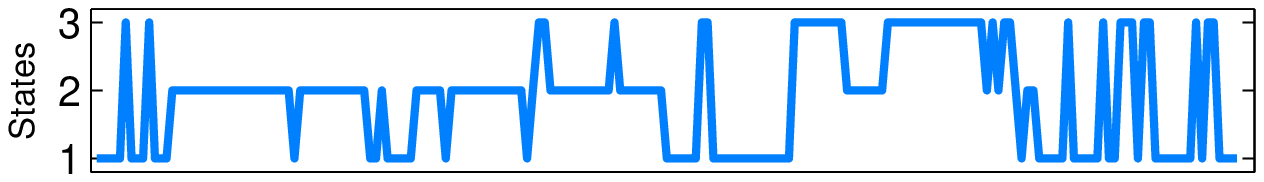}}
	\vspace{-0.5 cm}
	\end{minipage}
	
		\begin{minipage}[t]{1\linewidth}
		\centering
		\subfigure[(c)\label{Fig:SKS-states}]{\includegraphics[width=0.5\linewidth,keepaspectratio]{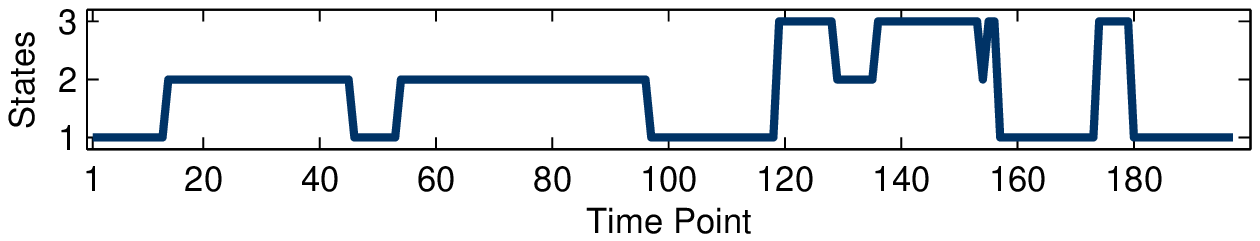}}
	\end{minipage}
\caption{Estimation of temporal dynamics of effective connectivity states in a real fMRI data for a subject. (a) fMRI ROI mean time-series. (b) Estimated factor time series using the f-SVAR model. (c) Estimated state sequence by the SKF (light blue) and SKS (dark blue). Dotted lines indicate regime segmentation by SKS.}
\label{Fig:fMRI-states-single-sub}
	\vspace{-0.2in}
\end{figure*}

To examine the transitions of the connectivity states in Figure \ref{Fig:state-net-mat} as a function of time, the estimated state-time alignment for the 10 subjects is shown in \ref{Fig:states_subs}. The results suggest that the effective connectivity states changes over time and the pattern of changes varied across subjects. However, the connectivity states reoccur over time and shared across subjects. It also exhibits slow dynamics, where the connectivity tends to be assigned to single discrete states for long periods, with occasional fast switching between states. Moreover, the degree of the non-stationarity differs between subjects, from the rapid transitions between states (subjects 3, 5 and 6) to almost time-constant connectivity remained in particular states, i.e state 3 (subjects 2, 4, 8), state 2 (subject 2) and state 1 (subject 9). Note that state 1 (yellow) with enhanced connectivity for all RSNs exhibits the lowest occurrence in the time-courses over all subjects. Figure \ref{Fig:fMRI-states-single-sub} shows the estimates for subject 6. The connectivity regime changes in the observed fMRI signals (Figure \ref{Fig:fMRI-states-single-sub}(a)) can be reflected in the lower-dimensional factor time series (Figure \ref{Fig:fMRI-states-single-sub}(b)). Besides, the SKS refines the state estimates by SKF, smoothing the spurious spikes and producing more stable regimes, as shown in (Figure \ref{Fig:fMRI-states-single-sub}(c)).

\clearpage
\section{Conclusion}

We developed a novel approach to identifying dynamic effective connectivity states with a large number of brain regions from fMRI data, based on a regime-switching factor model. The proposed approach first characterizes the high-dimensional fMRI data via a small number of factors for dimension reduction using a factor model in the observation space, and then performs connectivity regime segmentation in this low-dimensional latent factor subspace. By specifying the factor dynamics to follow a Markov-switching VAR process with a state-space formulation, it enables a reliable and efficient detection of change-points of the connectivity states using the Kalman smoothing and EM algorithm, and estimation of high-dimensional connectivity matrix for each state by projection of the estimated subspace parameters. The use of a regime-switching VAR specification allows us to examine state-driven changes in another important feature of connectivity, i.e. the directionality of connections, which are not addressed in earlier studies of dynamic functional (un-directional) connectivity. Hence, our approach provides a unified parametric framework for estimating both the time-varying connectivity structure and its quasi-stable state partitions, as distinct to using the separate steps of sliding-window connectivity analysis followed by K-means clustering. Moreover, the shortcomings of K-means clustering producing spurious fluctuations of states due to fixed-time windowing of time-varying connectivities and its failure to account for the temporal structure, can be overcome by the modeling with Markov chain which can capture both stable periods and the abrupt alternations of states via the transition probabilities. 

Simulation results demonstrate the superiority of our approach over the K-means clustering of TV-VAR coefficients, giving more accurate estimation of the dynamic states, and the within-state connectivity graph, particularly in the high-dimensional settings. In analyzing the resting-state fMRI data, the proposed estimator confirms previous findings of non-stationary, re-occurring brain states during rest, and state-dependent modulation of large-scale connectivity patterns and modular structure. Furthermore, we produced new evidence for across-state difference in both the strength and directionality of directed information flows within resting-state networks. Future works will investigate different variants of the proposed framework, e.g. by allowing regime-switching in the factor loadings, instead of the factor dynamics itself. The method can also be extended to analyze time-varying directed coherence which measures connectivity at specific frequency of brain activity.

\newpage

\bibliography{Ref-factor-SVAR}
\bibliographystyle{apalike}

\end{document}